\documentclass[11pt]{article} 
\newcommand{\myscale}{1} 

\usepackage{geometry}

\usepackage{natbib} 
\usepackage{graphicx}
\usepackage{microtype}
\usepackage{verbatim}
\usepackage{hyperref}
\usepackage{bm}
\usepackage{balance}
\usepackage{hyperref}
\hypersetup{pdfencoding=auto,unicode, psdextra, hidelinks} 
\usepackage[full]{textcomp}

\usepackage{url}
\usepackage[scaled=\myscale]{newtxtext}
\usepackage[scaled=\myscale, varg]{newtxmath}
\usepackage{xcolor, xspace}
\newcommand{\CC}{Clausius--Clapeyron\xspace}

\allowdisplaybreaks

\providecommand{\bmdefine}{}
\renewcommand{\bmdefine}[2]{\mycommand{#1}{{\bm{#2}}}}

\newcounter{regequation}
\newcounter{mycounter}
\newcommand{\newcm}[2]{\newcommand*{#1}{\ensuremath{#2}}}

\newcommand{\newc}{\newcommand*}
\newcommand{\mycommand}[2]{\providecommand{#1}{} \renewcommand{#1}{#2}}

\providecommand{\what}{\widehat}
\providecommand{\wtilde}{\widetilde}

\renewenvironment{quote}
 {\begin{list}{}
 {%\setlength{\leftmargin}{1cm}
 \setlength{\rightmargin}{\leftmargin}}
 \item[]}%
  {\end{list}}

  {\setcounter{equation}{\value{regequation}}
    \end{minipage}\end{shaded}\end{minipage}\end{center}\end{table} }

  {\setcounter{equation}{\value{regequation}}
   \end{minipage}\end{shaded}\end{minipage}\end{center}\end{table} }

 {\setcounter{equation}{\value{regequation}}
   \end{boxedminipage}\end{center}\end{table} }

    {\setcounter{equation}{\value{regequation}}
      \end{boxedminipage}\end{center}\end{table} }

   {\end{sffamily} \end{small}}

  {\newpage \end{list} \end{small} \end{sffamily}  }
   {\newpage \end{list} \end{small} \end{sffamily}}

{\end{list}\end{sffamily}\end{small}}

{\end{list}\end{sffamily}\end{small}}

 {\end{list}}

 {\end{list}}

 {\end{list}}

 {\end{list}}

\newcommand{\eqnab}%
    {}
\newcommand{\eqnabc}%
    {}
	\newcommand{\eqnc}%
	    {}
\newcommand{\eqnabbox}%
   {}

\newcommand{\aleqnab}%
	{}
	\newcommand{\aleqnbc}%
		{}
	\newcommand{\aleqnc}%
	{}
	\newcommand{\aleqnabc}%
		{}
		\newcommand{\aleqna}%
		{}

\newcommand{\itsf}{\itshape\sffamily}

\providecommand{\bneg}{\mkern -3mu} % negative space for subscripts after large brackets.
\providecommand{\bnegsup}{\bneg} % negative space for superscripts after large brackets.

\newcommand{\gobble}[1]{\!}

\newc{\x}{$x$\xspace} \newc{\y}{$y$\xspace}  \newc{\z}{$z$\xspace}

\newc{\probref}[1]{\thechapter.\ref{#1}}

\newc{\letrn}[2]{\lettrine[lines=3,lhang=0]{{\color{gray}#1}}
               {\MakeUppercase{\footnotesize\sffamily #2}}}
\newc{\figref}[1]{Fig.~\ref{#1}}
\newc{\Figref}[1]{Figure \ref{#1}}
\newc{\secref}[1]{section \ref{#1}}

\newc{\dbint}{\iint}

\newc{\curlz}{\mathrm{curl}_z}
\renewcommand{\div}[1][]{\nabla_{\! #1} \cdot}

\bmdefine{\omb}{\omega}
\bmdefine{\Omb}{\varOmega}

\mycommand{\wbtil}{\wtilde{\wb}}
\bmdefine{\taub}{\tau}
\bmdefine{\deltab}{\delta}
\newc{\ttaub}{\wtilde{\taub}}
\newc{\ttau}{\wtilde\tau}

\newc{\s}{\,\text{s}\xspace} 
\mycommand{\m}{\,\text{m}\xspace}
\newcommand{\J}{\,\text{J}\xspace}
\newc{\cm}{\,\text{cm}\xspace}
\newc{\km}{\,\text{km}\xspace}
\newc{\kg}{\,\text{kg}\xspace}
\newcm{\ps}{\,\text{s}{^{-1}}\xspace}
\newc{\mps}{\ensuremath{\m\ps}\xspace}
\newc{\Kv}{\,\text{K}\xspace}
\newc{\Sv}{\,\text{Sv}\xspace}
\newc{\hpa}{\,\text{hPa}\xspace}
\newc{\hPa}{\,\text{hPa}\xspace}
\newc{\rms}{\text{rms}}
\newc{\qg}{quasi-geostrophic\xspace}
\newc{\pg}{planetary-geostrophic\xspace}
\newc{\BV}{Brunt--V\"ais\"al\"a\xspace}

\newc{\Elnino}{El NI\~NO\xspace}
\newc{\crn}{\nonumber  \\}
\newc{\qqquad}{\qquad \qquad}
\newc{\cdel}{\cdot \nabla}

\newcommand{\eten}[1]{\ensuremath{\times 10^{#1}}\xspace}
\newc{\Ro}{\ensuremath{\mathit{Ro}}\xspace}

\newc{\Ek}{\ensuremath{\mathit{Ek}}\xspace}
\newc{\Ra}{\ensuremath{\mathit{Ra}}\xspace}
\newc{\Fr}{\ensuremath{\mathit{Fr}}\xspace}
\newc{\Bu}{\ensuremath{\mathit{Bu}}\xspace}
\newc{\Ri}{\ensuremath{\mathit{Ri}}\xspace}
\newc{\Rot}{\ensuremath{\mathit{Ro_\mathrm{t}}}\xspace}

\newc{\bpp}[2]{\bfrac{\partial #1}{\partial #2}}
\newc{\bppp}[2]{({\partial #1}/{\partial #2})}
\newc{\bfrac}[2]{\left( \frac{#1}{#2} \right) }
\newc{\textfrac}[2]{{(#1)}/{(#2)}}
\newc{\dbfrac}[2]{\left( \dfrac{#1}{#2} \right) }
\newc{\bfracs}[3]{\left(\frac{#1}{#2}\right)_{\bneg #3}}
\newc{\bfracsub}[3]{\left(\frac{#1}{#2}\right)_{\bneg #3}}
\newc{\bfracsup}[3]{\left(\frac{#1}{#2}\right)^{\bnegsup #3}}
\newc{\bpps}[3]{\left(\frac{\partial #1}{\partial #2}\right)_{\bneg #3}}
\newc{\bppsub}[3]{\left(\frac{\partial #1}{\partial #2}\right)_{\bneg #3}}
\newc{\bppsup}[3]{\left(\frac{\partial #1}{\partial #2}\right)^{\bnegsup #3}}
\newc{\bdds}[3]{\left(\frac{\d #1}{\d #2}\right)_{\bneg #3}}

\mycommand{\T}{\mathcal{T}}
\mycommand{\C}{\mathcal{C}}
\mycommand{\Eta}{\mathcal{E}}
\mycommand{\Zeta}{\mathcal{Z}}

\newc{\K}{\mathcal{K}}

\newc{\psisp}[1]{\psi^{(#1)}}

\newc{\psibar}{\overline \psi}
\newc{\Psibar}{\overline \Psi}
\newc{\phibar}{\overline \phi}
\newc{\thetabar}{\overline \theta}
\newc{\varthetabar}{\overline \vartheta}
\newc{\varphibar}{\overline \varphi}
\renewcommand{\hbar}{\overline h}

\newc{\qbar}{\overline q}
\newc{\ubar}{\overline u}
\newc{\vbar}{\overline v}

\newc{\ubbar}{\overline \ub}
\newc{\vbbar}{\overline \vb}
\newc{\bbar}{\overline b}
\newc{\pbar}{\overline p}
\newc{\zbar}{\overline z}
\newc{\rhobar}{\overline \rho}
\newc{\Fbar}{\overline F}
\newc{\Jbar}{\overline J}
\newc{\Sbar}{\overline S}
\newc{\Tbar}{\overline T}
\newc{\bybar}{\overline \by}

\newc{\zetabar}{\overline \zeta}
\newc{\etabar}{\overline \eta}
\newc{\betahat}{\ensuremath{\what \beta}}
\newc{\kbeta}{k_\beta}
\newc{\that}{{\what t}}
\newc{\khat}{{\what k}}

\newc{\deltahat}{\what \delta}
\newc{\etahat}{\what \eta}
\newc{\ubhat}{\what\ub}
\newc{\vbhat}{\what\vb}
\newc{\xhat}{{\what x}}
\newc{\xbhat}{{\what \xb}}
\newc{\yhat}{{\what y}}
\newc{\zhat}{{\what z}}
\newc{\uhat}{\what u}
\newc{\vhat}{\what v}

\newc{\bhat}{\what b}
\newc{\fbhat}{\what{\bm{f}}}
\newc{\phihat}{\what \phi}
\newc{\byhat}{\what \by}
\newc{\hhat}{\ensuremath{\what h}}
\newc{\fhat}{\ensuremath{\what f}}
\newc{\psihat}{\what \psi}
\newc{\Psihat}{\what \Psi}
\newc{\Nhat}{\what N}
\newc{\Nbar}{\overline N}

\newc{\qhat}{\what q}

\bmdefine{\jbi}{j}

\DeclareMathSymbol{\mywidetilde}{\mathord}{largesymbols}{"65}
\DeclareMathSymbol{\mywidehat}{\mathord}{largesymbols}{"62}

\renewcommand{\wbtil}{\mkern 4mu {\mywidetilde \mkern-14mu \wb \mkern 0mu}}

\newc{\tauhat}{\what \tau}

\mycommand{\Ebar}{\overline E}
\mycommand{\Zbar}{\overline Z}

\newc{\ug}{u_g}
\newc{\vg}{\ensuremath{v_g}\xspace}
\newc{\Lr}{\Ld}
\newcommand{\Ld}{\ensuremath{L_{d}}\xspace}

\newc{\lr}{\ensuremath{l_r}}
\newc{\lrr}{\ensuremath{\lambda_r}}
\newc{\vrt}{\vartheta}
\newc{\cross}{\times}
\newc{\half}{\frac 12}
\newc{\third}{\frac 13}
\newc{\quarter}{\frac 14}

\newc{\dt}{\, \mathrm{d}t}
\newc{\dx}{\, \mathrm{d}x}
\newc{\dy}{\, \mathrm{d}y}
\mycommand{\dz}{\, \mathrm{d}z}
\newc{\dk}{\, \mathrm{d}k}
\newc{\dr}{\, \mathrm{d}r}
\newc{\dkb}{\, \mathrm{d}\bm{k}}
\newc{\dxb}{\, \mathrm{d}\xb}
\newc{\dvrt}{\, \mathrm{d}\vrt}
\newc{\dpr}{\, \mathrm{d}p}

\newc{\dAb}{\, \mathrm{d} \bm{A}}
\newc{\Dth}{{\D_h \over \D t}}
\newc{\Dto}{{\D_0 \over \D \that}}
\newc{\Fb}{\bm{F} }
\bmdefine\Fb{F}

\newc\Ab{\ensuremath{\bm{A}}}
\newc\Cb{\ensuremath{\bm{C}}}
\newc\Jb{\ensuremath{\bm{J}}}
\newc\Bb{\ensuremath{\bm{B}}}
\newc\xb{{\bm{x}}}
\bmdefine{\pb}{p}
\bmdefine{\qb}{q}
\bmdefine{\ab}{a}
\bmdefine{\Xb}{X}
\bmdefine{\Yb}{Y}
\bmdefine{\rb}{r}

\bmdefine{\sb}{s}
\bmdefine{\Db}{D}

\bmdefine{\kbi}{k}
\bmdefine{\jbi}{j}
\bmdefine{\Vb}{V}
\bmdefine{\Wb}{W}

\newc{\fb}{{\ensuremath{\bm {f}}}}
\newc{\fbo}{{\ensuremath{\bm {f}_0}}}
\def\vb{{\ensuremath{\bm {v}}}\xspace}
\def\ub{{\ensuremath{\bm {u}}}\xspace}
\def\wb{{\ensuremath{\bm {w}}}\xspace}

\bmdefine{\Ub}{U}
\bmdefine{\cb}{c}
\bmdefine{\psib}{\psi}
\bmdefine{\phib}{\phi}

\newcommand{\by}{\ensuremath{b}}

\newc{\Tb}{{\bm T} }
\newc{\lhs}{left-hands side\xspace}
\newc{\fo}{\ensuremath{f_0}\xspace}
\def\DD#1{{\D#1 \over \D t}}

\newc{\MD}[1]{{\partial {#1} \over \partial t} + (\vb \cdot \del) #1}
 
\newcommand{\pp}[3][]{\frac{\partial^{#1} #2}{\partial #3^{#1}}}

\newc{\ppa}[1]{{\partial \over \partial #1}}
\newc{\pt}{{\partial \over \partial t}}
\newc{\dell}[1][]{\nabla_{\!#1\,}}
\newc{\del}{\nabla}
\newc{\grad}{\nabla}
\newc{\delh}{\nabla_h}
\newc{\cp}{c_{p}}
\newc{\cv}{c_{v}}

\renewcommand{\d}{\mathrm{d}}

\newcommand{\D}{\mathrm{D}}

\newcommand{\textmonth}{\ifcase\month \or January \or February\or
March\or April\or May \or June \or July\or August\or September\or October
\or November\or December\fi\xspace}

\DeclareMathAlphabet{\mathsfsl}{\encodingdefault}{hlst}{b}{sl}   % uses lucida
\newc{\putabcd}{\put(-9.2,7){\small \textit{\textsf{(a)}}}
\put(-4.5,7){\small \textit{\textsf{(b)}}}
\put(-9.2,3.3){\small \textit{\textsf{(c)}}}
\put(-4.5,3.3){\small \textit{\textsf{(d)}}}
}

\newc{\putab}{\put(-8.9,3.3){\small \textit{\textsf{(a)}}}
\put(-4.3,3.3){\small \textit{\textsf{(b)}}}
}

\bmdefine{\Ebc}{\mathcal E}
\newcommand{\bfig}{\begin{figure}}
\newcommand{\efig}{\end{figure}}
\newcommand{\bcen}{\begin{center}}
\newcommand{\ecen}{\end{center}}
\newcommand{\beq}{\begin{equation}}
\newcommand{\eeq}{\end{equation}}

\newcommand{\bsub}{\begin{subequations}}
\newcommand{\esub}{\end{subequations}}

\AtBeginDocument{}
 
\newcommand{\Hc}{\mathcal H}
\newcommand{\CAPE}{\textsc{cape}\xspace}

\newcommand{\MG}{Matsuno--Gill\xspace}
\newcommand{\qsat}{q^*} 
\newcommand{\alphahat}{\widehat\alpha}
\usepackage{float}
\usepackage{afterpage}

\newcommand{\figwidth}{\textwidth}
\newcommand{\figheight}{4.6cm}
\newlength{\colwidth}
\title{Convective Organization and Eastward Propagating Equatorial Disturbances in a Simple Excitable System}  
\author{Geoffrey K Vallis and James Penn\\[0.1cm] University of Exeter}
\date{\footnotesize 23 August, 2019}

\begin{document}

\renewcommand{\div}{\nabla \cdot}
\maketitle

\begin{abstract} %\normalsize
We describe and illustrate a mechanism whereby convective aggregation and eastward propagating equatorial disturbances, similar in some respects to the Madden--Julian oscillation,  arise.  We construct a simple, explicit system consisting only of the shallow water equations plus a humidity variable; moisture enters via evaporation from a wet surface, is transported by the flow and removed by condensation, so providing a mass source to the height field.   For a broad range of parameters the system is excitable and self-sustaining, even if linearly stable, with condensation producing convergence and gravity waves that, acting together, trigger more condensation.  On the equatorial beta-plane the convection first aggregates near the equator, generating patterns related to those in the Matsuno--Gill problem. However, the pattern is unsteady and more convection is triggered on its eastern edge, leading to a precipitating disturbance that progresses eastward.  The effect is enhanced by westward prevailing winds that increase the evaporation east of the disturbance.  The pattern is confined to a region within a few deformation radii of equator because here the convection can best create the convergence needed to organize into a self-sustaining pattern.   Formation of the disturbance preferentially occurs where the surface is warmer and sufficient time (a few tens of days) must pass before conditions arise that enable the disturbance to reform, as is characteristic both of excitable systems and the MJO itself.  The speed of the disturbance depends on the efficiency of evaporation and the heat released by condensation, and is typically a few meters per second, much less than the Kelvin wave speed. 

\end{abstract}

%\section{Introduction}

\section{Introduction}
{T}wo, related, problems in tropical dynamics concern convective organization and the Madden--Julian oscillation (the MJO), both of which have been the subject of considerable investigation. However, no consensus has been reached in either case as to the key mechanisms involved and indeed, as recently as 2017,  \citet{Fuchs_Raymond17} wrote that `the MJO [has] been and still remains a ``holy grail'' of today’s atmospheric science research'.

The MJO is a large-scale precipitating disturbance that propagates eastward at a few meters per second, centered near the equator and extending meridionally about 20\textdegree\ North and South \citep{Madden_Julian71, Zhang05, Lau_Waliser12}. Its influence in the geopotential field extends zonally many thousand kilometers,  and although the  region of intense precipitation  is confined to the tropics its influence can be felt in the mid-latitudes and stratosphere.  Typically, the disturbance forms over the warm waters of the Indian Ocean and progresses east over the maritime continent and across the Pacific, decaying over the cooler waters of the eastern Pacific. The MJO reforms with a timescale of order a few tens of days -- it  is sometimes called the 30--60 day oscillation --  and its spectral signature is manifest at a lower frequency than a Kelvin wave of the same scale \citep{Wheeler_Kiladis99, Kiladis_etal09}.  The MJO may also be thought of as a translating disturbance that recurs on the above timescale, and that is the perspective we will take here. 

The eastward propagation of the MJO suggests the influence of Kelvin waves, yet its propagation speed is much less than that of dry Kelvin waves (about 20\mps) for any reasonable equivalent depth of the atmosphere. The condensation of water vapor is likely to be a significant influence on the propagation of any precipitating disturbance and this led to theories centered around the notion of a `moisture mode' \citep[e.g.,][]{ Raymond01, Raymond_Fuchs09,  Sobel_Maloney13}. The framework of \cite{Majda_Stechmann09} also involves moisture in an essential way.  Although the various models differ from each other in their assumptions and dynamics, a commonality is to assume some low-mode structure and calculate a dispersion relation or mode of instability that, depending on the form of the model, gives rise to  propagation. Rather different types of model were presented by \cite{Biello_Majda05} and \cite{Yang_Ingersoll13}, who suggested that the MJO is a multi-scale phenomenon with the convective structures resulting from a smaller scale activity. 

In contrast to these relatively simple and/or semi-analytic models, comprehensive three-dimensional models  are becoming able to simulate many aspects of the MJO \citep[e.g.,][]{Liu_etal09, Arnold_Randall15, Khairoutdinov_Emanuel18}.   In these simulations the MJO seems to be associated with some form of instability associated with a very large scale convective aggregation.  However, the complexity of the models can make it difficult to determine the dominant mechanism and to make connections with or distinguish between the multi-scale theories and large-scale moisture-mode models.

Even without reference to the MJO, convective aggregation \citep[e.g.,][]{Bretherton_Blossey05,  Muller_Bony15,  Wing_etal17} is a fraught subject with a diversity of results across models and no single accepted dominant mechanism. `Self'-aggregation, meaning the aggregation of convection over a surface of uniform temperature with no external large-scale influences, may be particularly sensitive to the parameters of the situation. However, as we will discuss, organization in a differentially rotating frame (essentially the beta plane) may be much more robust and is likely to be the building block of the MJO.

In this paper we explore these issues and identify a robust mechanism of equatorial aggregation and eastward propagation. We do so through the use of a minimal but explicit model, specifically the moist shallow water equations with relatively simple physics, a system that retains many of the properties of the three-dimensional equations.  We try to make no assumptions that are not transparently connected to the properties of the equations of motions themselves. We find that the resulting system is an excitable one that, even when linearly stable, can produce self-sustained, irregular motion that, when solved on a beta-plane, gives rise to convective organization and eastward propagating equatorial disturbances with many similarities to the MJO. More generally, by considering tropical convection as an excitable system, properties that may have appeared surprising then seem less so, even where not fully understood. 

We first describe the equations themselves. We then show the system admits of exact solution of no motion, and examine the stability and excitability properties of the system. After that we describe a number of numerical simulations and discuss the mechanisms of organization and eastward propagation. In the final section we place our results in a broader context and provide some conclusions. 

\section{The Moist Shallow Water Equations} \label{sec:msweq}

The shallow water equations may be derived by way of an expansion of the hydrostatic primitive equations in terms of vertical modes \citep{Matsuno66, Gill80, Vallis17}. The first baroclinic mode then obeys an equation set similar to the reduced-gravity shallow water equations although with slight differences in interpretation -- in particular the height field is related to the temperature field, and a heating is represented as a source of sink of mass.   The usual derivation is valid only in the linear case, since the nonlinear terms bring in additional vertical modes, but nonetheless equations of this general form are often regarded as good analogs of the primitive equations in cases when the vertical structure is dominated by the first baroclinic mode \citep[e.g.,][]{Yano_etal95, Sobel_etal01, Majda_Stechmann09}. A deep baroclinic mode is also an observed signature of the MJO \citep{Kiladis_etal05, Adames_Wallace14}. 

We include moisture by introducing  humidity variable, $q$,  that is carried by the fluid and that is conserved in the absence of evaporation and condensation, and when condensation occurs it provides a heat source that affects the height field.  Since moisture has a much smaller scale height than the troposphere  (about 2\km rather than 8\km) we take the velocity field to be that of the lower atmosphere.  The equations of motion then become, written using standard shallow water notation and omitting frictional and diffusional terms,
\begin{subequations}
\label{msw.1}
\begin{align}
    \DD \ub + \fb \times \ub & = - g  \del h,  \\
    \pp h t +  \div ( h \ub )& =  - \gamma C + R, \\
    \pp q t + \div (q \ub ) & = E - C,
\end{align}
\end{subequations}
where $E$ represents evaporation, $C$ represents condensation, $R$ represents thermal forcing (e.g., a radiative relaxation) and $\gamma$ is a parameter proportional  to the latent heat of condensation.   With no evaporative or thermal relaxation terms, but even in the presence of condensation,  (\ref{msw.1}b) and (\ref{msw.1}c) combine to give the conservation equation
\begin{equation}
	\label{msw.2} 
   	\pp M t +  \div ( M \ub ) = 0, 
\end{equation}
where $M = h - \gamma q$ is an analog of a moist static energy or moist enthalpy for the system.   The above moist shallow water (MSW) equations are similar to those used by \cite{Bouchut_etal09} and \cite{Rostami_Zeitlin18}, although the implementation and the experiments, described below, significantly differ. 

Evaporation from a wet surface is parameterized by a bulk-aerodynamic-type formula,
\begin{equation}
	\label{msw.3} 
   	E = \lambda |\ub/U_0| (q_g - q )  \Hc (q_g -  q),    
\end{equation}
where $\lambda$ is a constant (similar to a drag coefficient) and $q_g$ is the surface humidity.  The Heaviside function, $\Hc$,  ensures that evaporation only occurs when the surface humidity is larger than that of the atmosphere, and dew formation is forbidden. The dependence on velocity ($|\ub|/U_0$, where $U_0$ is a scaling constant) enables wind-induced evaporative effects to occur, and may be omitted. 

Condensation is allowed to occur on saturation and we take it to be 'fast', meaning that it occurs on a timescale faster than any other in the system and does not allow the fluid to become significantly supersaturated. We represent this as 
\begin{equation}
	\label{msw.4} 
    C  = \Hc (q - \qsat) { (q - \qsat) \over \tau }  ,  %\qquad  q_\text{sat} = q_0 \exp(-\alpha h)
\end{equation}
where $\qsat$ is the saturation specific humidity and $\tau$ is the timescale of condensation. (Fast condensation is a simplification of `fast autoconversion', commonly used in cloud resolving models.)    Precipitation schemes of this form are a common feature of various idealized GCMs \citep[e.g.,][]{Frierson_etal06} and simple (e.g., Betts--Miller \cite{Betts86} and convective-adjustment style) convection schemes. More elaborate convection parameterization schemes do exist for the shallow water equations \citep{Wursch_Craig14} but here we try to minimize their use.  An expression for $\qsat$ in terms of $h$ is derived in the next section. 

%Over the range of temperatures commonly encountered in Earth's atmosphere the saturation vapor pressure increases approximately exponentially with temperature and we represent this by
%\begin{equation}
%	\label{msw.5} 
%   	 \qsat = q_0 \exp(-\alpha h/H) , 
%\end{equation}
%where $q_0$, $\alpha$ and $H$ are constants, and  the moisture is assumed to be removed from the system with no re-evaporation.  

Equations \eqref{msw.1}, \eqref{msw.3} and \eqref{msw.4} form a complete set. We now discuss the form of equations that we numerically integrate  and the values of the parameters in relation to the true atmosphere.

\subsection{Semi-Linear Equations of Motion}

We will use the linear form of the momentum and height equations, but keep the  nonlinearity in the moisture and \CC equations.  The reason for this is that our main focus is on equatorial dynamics and the linear shallow water equations are remarkably fecund source of knowledge in that arena, beginning with \cite{Matsuno66} and \cite{Gill80}.  Tropical cyclones  are also (deliberately) eliminated by such a linearization.   However, a comparable linearization of the moisture equation, about either an unsaturated or saturated state, would be a poor approximation because of the large range of moisture values and the spatial variation of $\qsat$, and convective triggering by a cold gravity wave be misrepresented. Since $q$ and $\qsat$ do vary, it is not the case that moisture convergence alone leads to a convective instability.  

The equations of motion become
\begin{subequations}
\label{eqp.1}
\begin{align}
    \pp \ub t + \fb \times \ub & = - g \del h + \nu_u \del^2 \ub ,  \\
    \pp h t +  H \div \ub  & =  - \gamma C +  {(h_0 - h) \over \tau_r } + \nu_h \del^2 h, \\
    \pp q t + \div (q \ub ) & = E - C + \nu_q \del^2 q ,
\end{align}
\end{subequations}
where $H$ is the equivalent depth of the basic state and $h$ is now the deviation from this.  (Only the product of $g$ and $H$  matter but we retain familiar shallow water notation.)  The term $(h -  h_0)/\tau_r$ represents a radiative relaxation back to a prescribed value $h_0$ (usually taken to be zero) on a timescale $\tau_r$, and we now explicitly include diffusive terms with obvious notation.

\subsubsection{Parameters and Relation to Atmosphere}
Temperature variations may be roughly related to height variations by $h/H \sim -\Delta T /T_0$ where $T_0$ is a constant (e.g., 300 K), and we choose $H$ = 30\m and $g = 10\m\s^{-2}$ \citep[e.g.,][]{Kiladis_etal09}. An appropriate radiative relaxation timescale  $\tau_r$ is of order a few days, appropriate for a lower atmosphere \citep[e.g.,][]{Gill80}. We do not include  drag on velocity in the simulations shown here, but simulations with a moderate drag are similar. The  coefficient of evaporation $\lambda$ is such that for a typical velocity (which is of order 1\mps), or with no velocity dependence, the system becomes saturated in a few days. 

Now consider the other moist parameters.  In an ideal gas the latent heat of condensation, $L$, is such that $L \Delta q_a = \cp \Delta T$ where $q_a$ is the specific humidity in the gas and $\cp$ is the heat capacity at constant pressure. Here the analogous relation is $\gamma \Delta q = - h$ and so we estimate $\gamma$ by
\begin{equation}
	\label{eqp.4} 
   	 \gamma \sim  \frac {L  H Q_a} {q_0 \cp T_0} \sim 8.5 \m, 
\end{equation} 
where $Q_a$ is a typical saturated value of specific humidity near the surface in the tropical atmosphere, $q_0$ is the corresponding quantity in the MSW equations, which  (without loss of generality) we take to be unity.  Using $L = 2.4 \eten 6  \J \kg^{-1} $,  $\cp = 1004 \J \kg^{-1} \Kv^{-1}$,  $Q_a = 0.035$, $T_0 = 300$ and $H = 30\m$ gives the value above. 

To obtain an appropriate expression for saturation humidity for use in \eqref{msw.4} we begin with the approximate solution of the \CC equation, namely
\begin{equation}
	\label{eqp.2} 
    e_s = e_0 \exp\left[\frac{L}{R^v} \left(\frac{1}{T_0} - \frac 1 T \right) \right] 
    \approx e_0 \exp\left[\frac{L \Delta T}{R^v T_0^2}\right]  , 
\end{equation}
where $e_s$ is the saturation vapor pressure, $e_0$ is a constant, $R^v$ is the gas constant for water vapor, and $\Delta T = T - T_0$.  Now,  since the saturation specific humidity varies approximately in the same way as the saturation vapor pressure, we can write an expression for the saturation humidity in our system as
\begin{equation}
	\label{eqp.3} 
    \qsat = q_0 \exp(-\alpha h/H), 
\end{equation}
where $\alpha\sim L/(R^v T_0)$.  With $R^v = 462 \,J\kg^{-1}\Kv^{-1}$ and $T_0 = 250\Kv$ we have $\alpha \sim  20$. Rather larger values are arguably more realistic since the shallow water equations mainly represent horizontal variations of temperature.  The diffusive parameters ($\nu_u$ etc.) are chosen on a numerical  basis.  All these parameter estimates are manifestly approximate, and variations of up to an order of magnitude may be reasonably explored. 

The above MSW equations thus include the effects of evaporation, transport and precipitation of moisture, the heat release associated with precipitation, and a temperature dependent saturation mixing ratio -- some of the main features of a tropical atmosphere.  Tropical convection is often described as having quasi-equilibrium nature on sufficiently long time scales \citep[e.g.,][]{Betts73, Arakawa_Schubert74, Emanuel_etal87}, associated with the convection occurring on a much faster timescale than that of the larger scale flow.  The production of convective available potential energy (\CAPE) by large-scale flow is then nearly balanced by the relaxation of \CAPE by convection and the temperature profile is constrained to be close to neutrally stable.  These effects are represented in our model by the use of a prescribed vertical structure and by condensational effects acting on a fast timescale. Convective instability will result if the fluid is near to saturation and the large-scale flow is convergent at the same location. However, convergence alone does not necessarily lead to convection, nor is convergence the only way to excite it; gravity waves can also trigger convection (as noted by \cite{Yang_Ingersoll13}), given the height dependence of the Clausius--Clapeyron equation.

\section{Exact Steady Solutions}

The equations of motion admit of steady solutions with no motion. In this state the height and moisture equations are then,
\begin{subequations}
\label{ss.1}
\begin{align}
    0  & =  - \gamma C  - \lambda_r h , \\
    0  & =    \lambda (q_g - q )  \Hc (q_g -  q)    -   C ,
\end{align}
\end{subequations}
where $\lambda_r = 1/\tau_r$, $C$ is given by \eqref{msw.4} and the saturation humidity is given by \eqref{eqp.3}. (We omit the velocity dependence on the evaporation,  but restore this in later sections.)

Suppose first that $q_0 \geq q_g$, which we might interpret as being the case with an unsaturated surface. In this case there can only be evaporation if the fluid is sufficiently cold and $h > 0$. However, the radiative forcing (and the effects of any condensation) will both warm the fluid and (\ref{ss.1}a) cannot be satisfied since both terms are negative. The solution in this case is then $h = 0$, $q = q_q$, $C=0$ and, because $q = q_g$, the evaporation is zero.  The fluid is thus not saturated unless $q_g = q_0$. 

Now suppose $q_g \geq q_0$. A balance can now occur when the evaporation equals the condensation and the radiative cooling equals the latent heat release. Eliminating the condensation term in (\ref{ss.1}a) and (\ref{ss.1}b) gives
\begin{equation}
	\label{ss.2} 
   	0 = - \lambda_r h - \gamma \lambda (q_g - q )   \Hc (q_g -  q) .
\end{equation}
Now, the assumed rapidity of the condensation means that the fluid is, to a very good approximation (asymptotically to $O(\tau)$) saturated and \eqref{ss.2} becomes
\begin{equation}
	\label{ss.3} 
   	0 = - \lambda_r h - \gamma \lambda \left(q_g - q_0 \exp(-\alphahat h) \right) ,
\end{equation}
where $\alphahat = \alpha/H$ and we omit the Heaviside term since $q_g \geq q$, but retain $q_0$ for clarity. If $q_g = q_0$ the solution is simply  $h = 0$ and $q = q_g$. If $q_g > q_0$ the exact solution of this equation may be written in terms of the Lambert $W$ function, namely the function that satisfies the equation $W(z) \exp(W(z)) = z $ for any $z$ \citep{Corless_etal96}, with  $h$ then given by
\begin{equation}
	\label{ss.4} 
   	h = - R  + \
       \frac 1 \alphahat W\left( {\alphahat \exp( { \alpha  R} )\over A }  \right),
\end{equation}
where $A = \lambda_r/(\gamma \lambda q_0)$ and  $R = q_g \gamma \lambda /\lambda_r$.   This solution is the shallow water analog of the drizzle solution in the vertically continuous problem found by \cite{Vallis_etal19}, and the precipitation in both cases is non-convective, involving no fluid motion.

A more easily interpreted expression results if we suppose that the saturation humidity varies linearly with the height field and $\qsat = q_0 (1 - \alphahat h)$. Equation \eqref{ss.3} becomes
\begin{equation}
	\label{ss.5} 
   0 = - \lambda_r h - \gamma \lambda \left(q_g - q_0 (1 - \alphahat h) \right)
\end{equation}
with solution
\begin{equation}
	\label{ss.6} 
 h =  \frac{  - \gamma \lambda (q_g - q_0) }  { \lambda_r + \gamma \lambda q_0 \alphahat}  ,
\end{equation}
and the humidity is the saturated value occurring at this value.  If $q_g = q_0$ then  $h = 0$, as expected, giving a solution with no evaporation or precipitation.  If $q_g > q_0$ then evaporation occurs, warming the fluid and reducing the value of $h$. The humidity, $q$, is then less than that at the surface, $q_g$, and evaporation is continuous.  In the steady state the resulting value of  $h$ is such that the evaporation exactly replenishes the condensation, and the condensational heating is exactly balanced by the radiative cooling.  For $q_g = 1.1 q_0$ and the other parameters taking the values previously derived, \eqref{ss.6} gives values of $h$ of order 0.1\m or less, considerably smaller than the variations we will find when the model is in a convective regime. 

As much their particular form, the expression \eqref{ss.4} and its approximation \eqref{ss.6} are important because they demonstrate that steady solutions of no motion do exist to the problem. We now examine whether these solutions are stable to small perturbations and whether the system is excitable.

\section{Stability and Excitability}
     \newcommand*{\putabc}{\put(-270,335){\small (a)}
     \put(-10,335){\small (b) }   \put(-200,235){\small (c) }
     }

 \afterpage{
\begin{figure}[H]
  %  \centering
    (a) \hspace{8cm} (b) \\
    \includegraphics[width=0.44\figwidth]{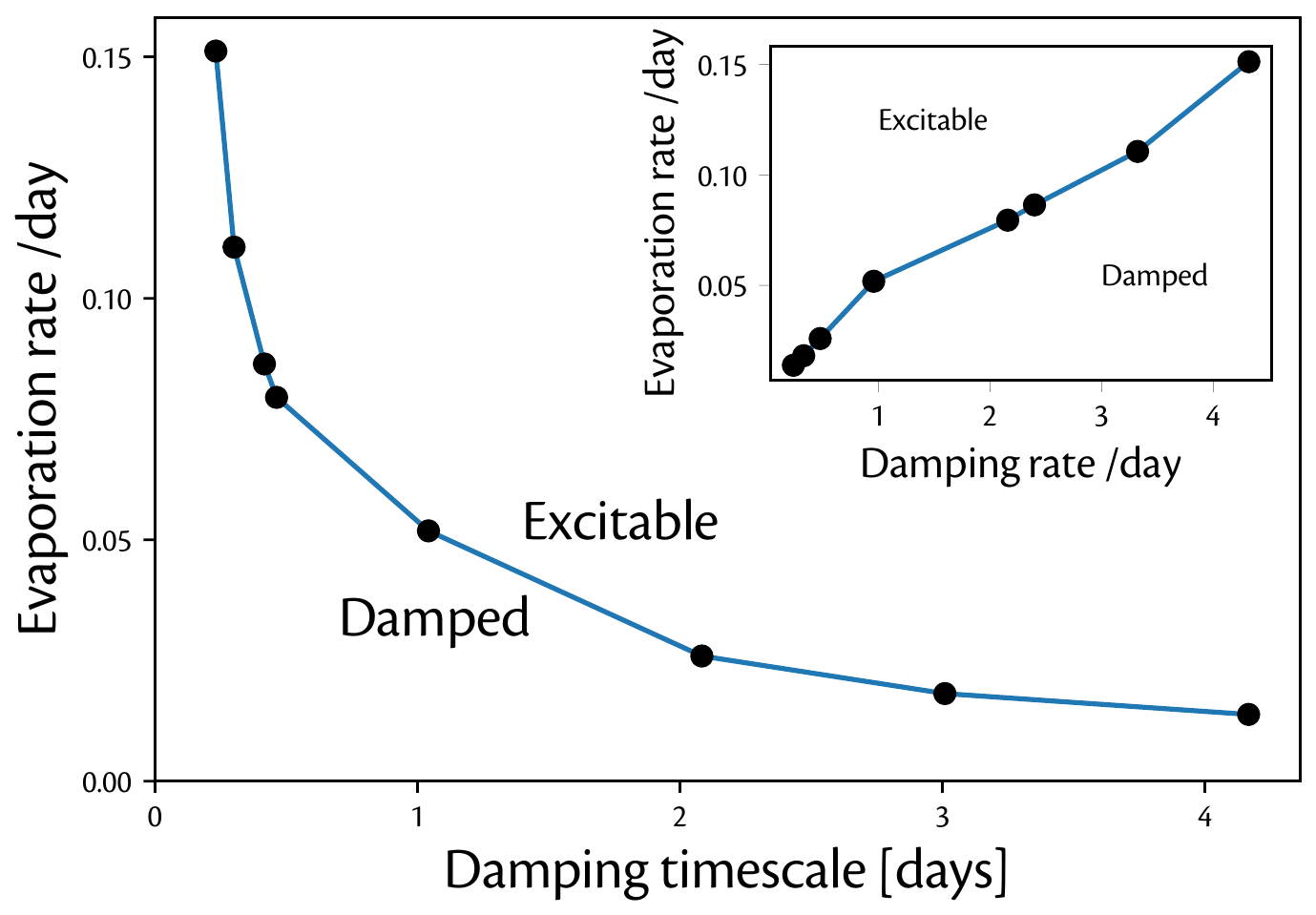} \qquad \qquad
         \includegraphics[width=0.41\figwidth]{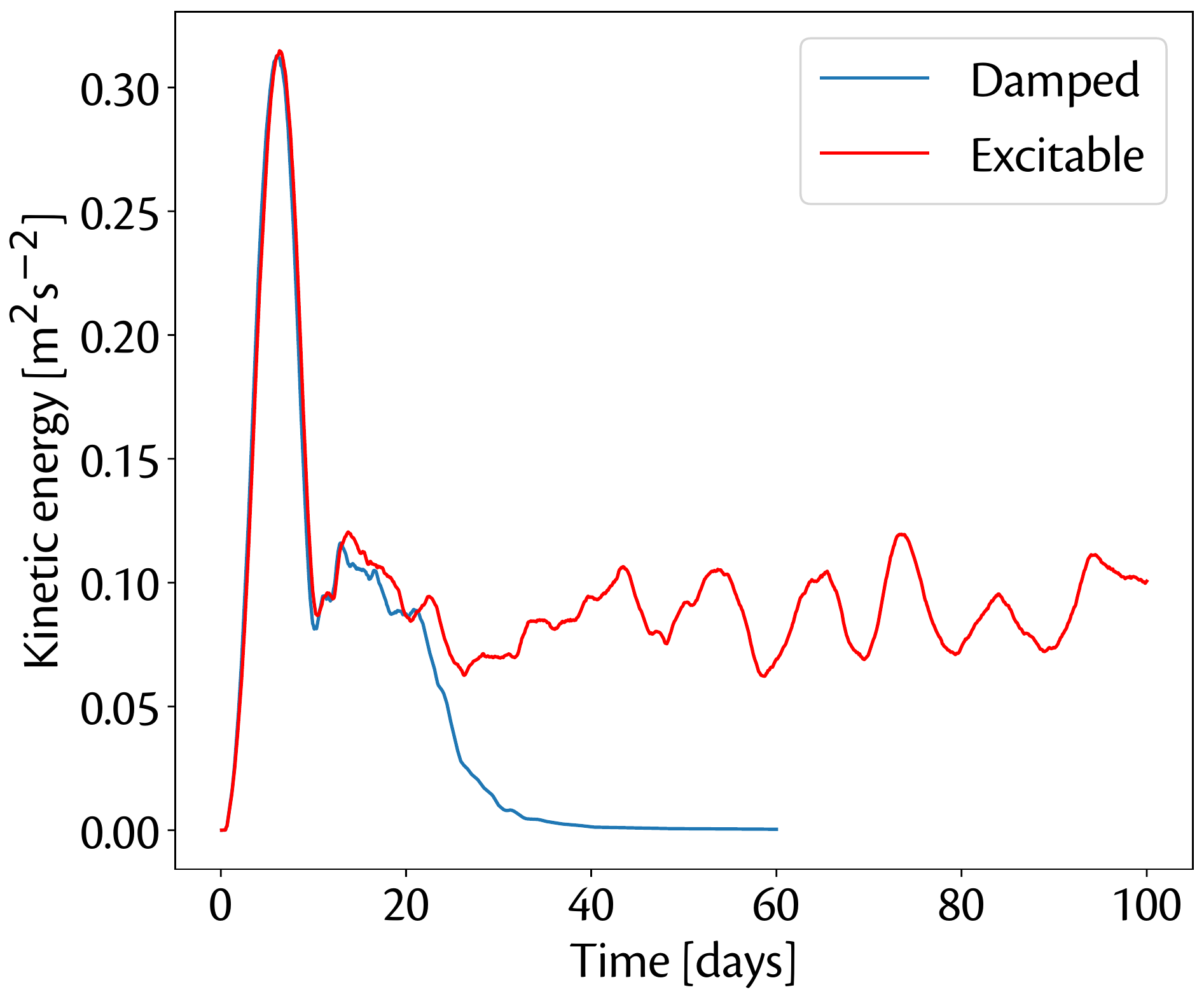}  \\ \hspace*{4cm} (c) \\
     \centerline{\includegraphics[width=0.41\figwidth, height=\figheight]{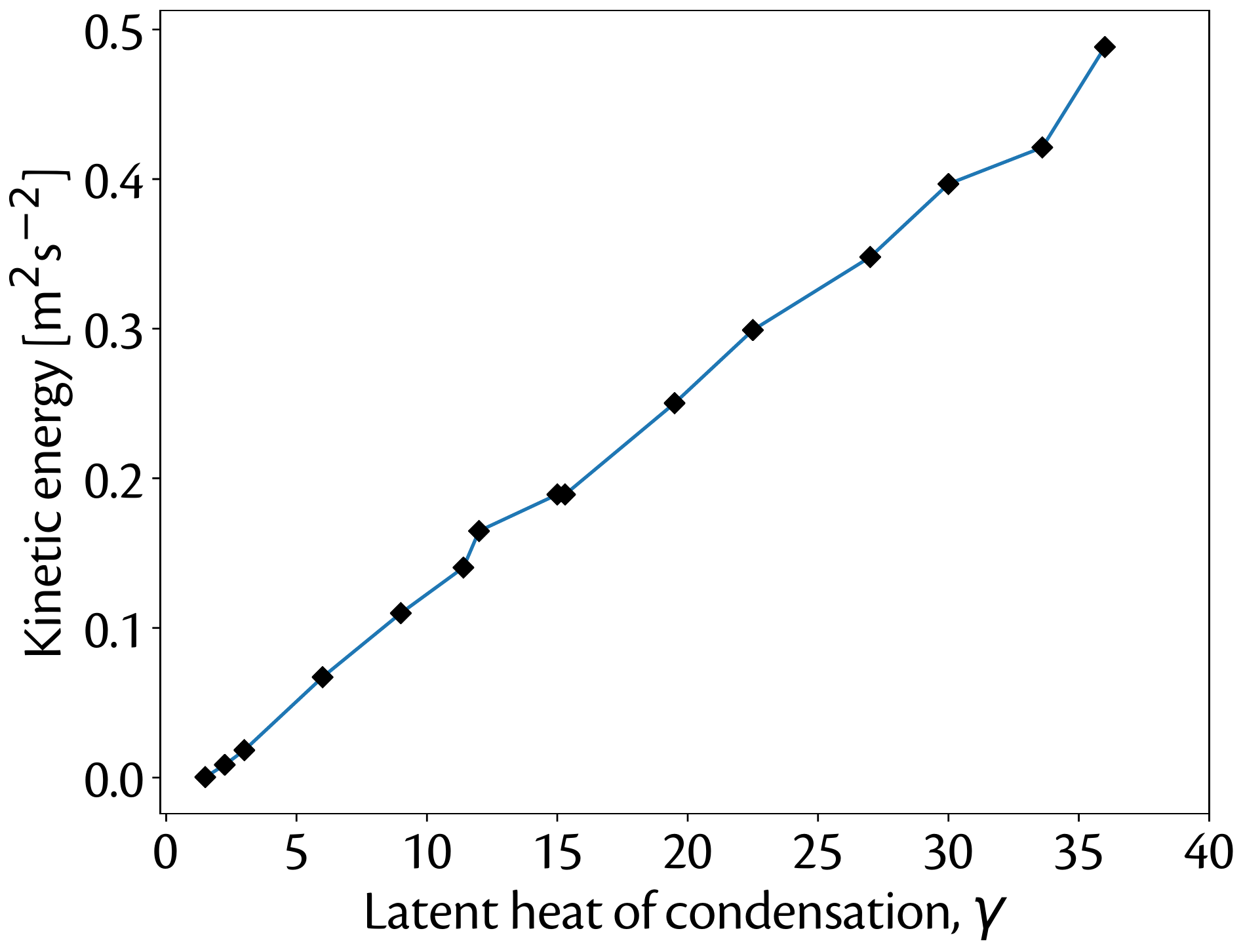} }
  %   \putabc
    \caption{a) Boundary of the excitable state for varying values of the radiative damping of the height field and the efficiency of evaporation. The stationary state is saturated with no precipitation of evaporation. In the excitable regime the system maintains a chaotic, convectively active state. In the stable state the system eventually returns to its original state of no motion, even though that state is linearly unstable.  (b) Evolution in an excitable and a damped states, both with a damping timescale of 2 days and with evaporative parameters on either side of the critical line.  (c) The domain and time-averaged kinetic energy as a function of the latent heat parameter, $\gamma$. }
    \label{fig:crit}
\end{figure}  
}

Unless the diffusive and damping terms in the equations are very strong, the solutions found above can be expected to be conditionally unstable, as can be seen by the following argument.  In the absence of motion the atmospheric humidity will be relax to the state $q = q_g$, and if $q_0 \leq q_g$ then the atmosphere is saturated, satisfying the steady solution calculated above. However, a localized perturbation, even an infinitesimal one,  will lead to further condensation and thence a perturbation in the height equation and the generation of gravity waves, as well as low-level convergence at the source of condensation. These gravity waves will propagate and will induce more condensation nearby, which will in turn generate more gravity waves and so on. Evidently, the steady saturated solution is unstable to an infinitesimal perturbation.  If, on the other hand, $q_0 >q_g$ then the steady solution will not be saturated and an infinitesimal perturbation will have little effect, but a perturbation large enough to trigger condensation will generate further motion.  Thus, unless the damping is unrealistically large any perturbation that causes saturation will lead to the generation of motion and a spreading field of precipitation as the gravity waves propagate. 

Following a perturbation, and depending on (a) the rate of evaporation from the surface, (b) the magnitude of the latent heat of condensation, and (c) the size of the damping terms, the system will either evolve into a sustained convective state,  or the perturbations will eventually die and the system will return to a state of no motion.    The return can occur, even if the initial state is linearly unstable, because  the precipitation occurs at a lower temperature than that of the drizzle solution, leaving the atmosphere unsaturated and non-precipitating as it slowly relaxes back to the initial state.  The moisture is slowly replenished by surface evaporation, but if that occurs on a timescale that is long compared that on which the gravity waves are damped then the atmosphere only reaches saturation when all the perturbations have died.  An external perturbation could nevertheless excite the system again. 

 On the other hand, with sufficiently small damping and sufficient latent heat release the system will evolve to state of self-sustained motion; that is to say, the system will be `excitable'.  Various forms of excitable system exist (e.g., networks of neurons, auto-catalytic chemical reactions, certain cellular automata) and there is no universal definition, but typically they are extended non-equilibrium systems that may have a linearly stable fixed point but that nevertheless are susceptible to finite perturbations \citep{Meron92, Izhikevich07}. In our system, the return or otherwise to the initial state does not sensitively depend on the linear stability properties of the initial state: numerical experiments show that self-sustained states of convection may exist even if $q_0 > q_g$, in which case the steady solution is not saturated and thus linearly stable, and the sustained convection is then subcritical. Conversely, if the damping is moderately large, the system may return to its initial state, and stay there, even if that state is saturated and linearly unstable.

 \afterpage{
 \begin{figure}[H] \centering 
     \includegraphics[width=\colwidth]{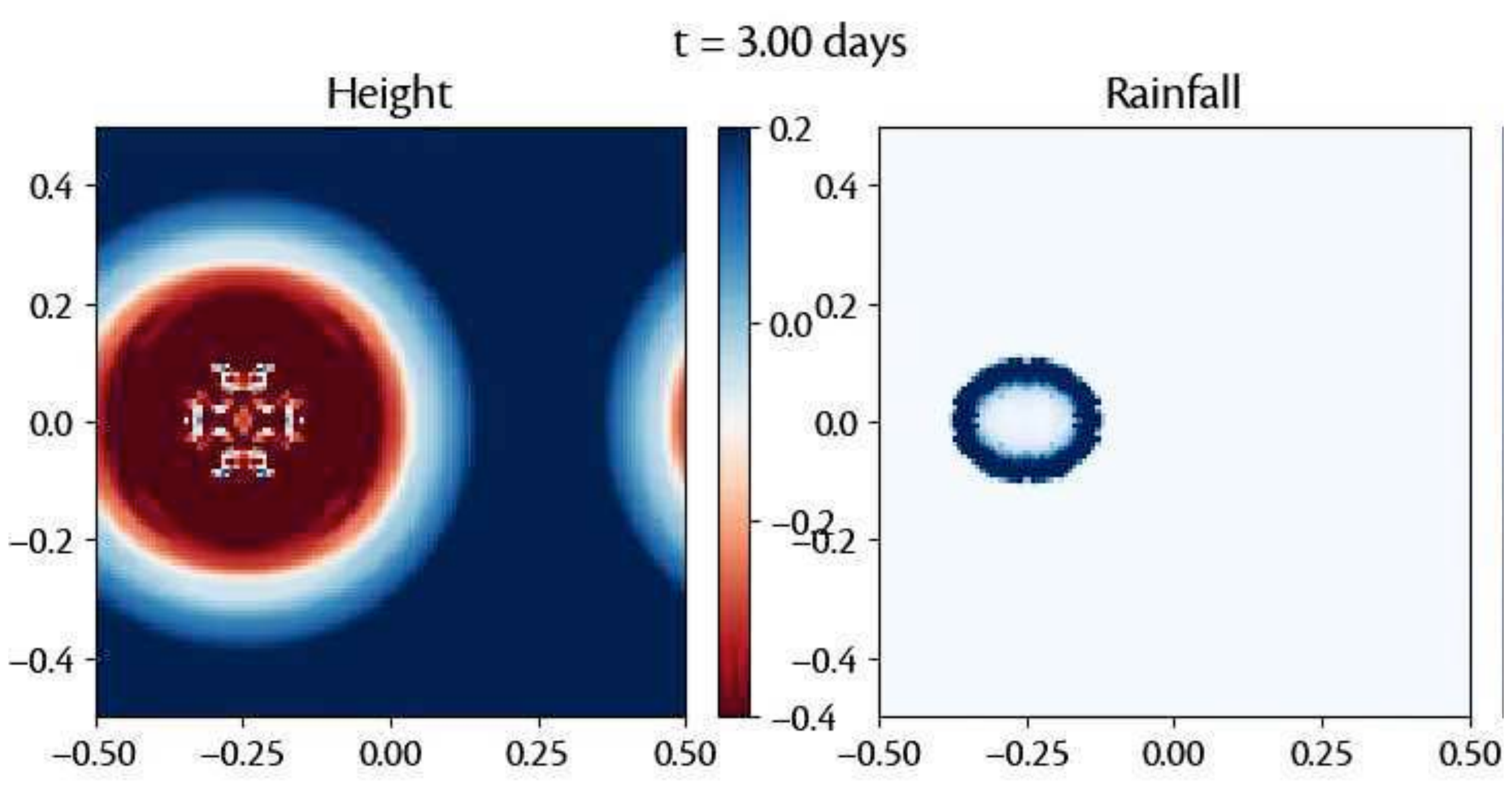} \\
     \includegraphics[width=\colwidth]{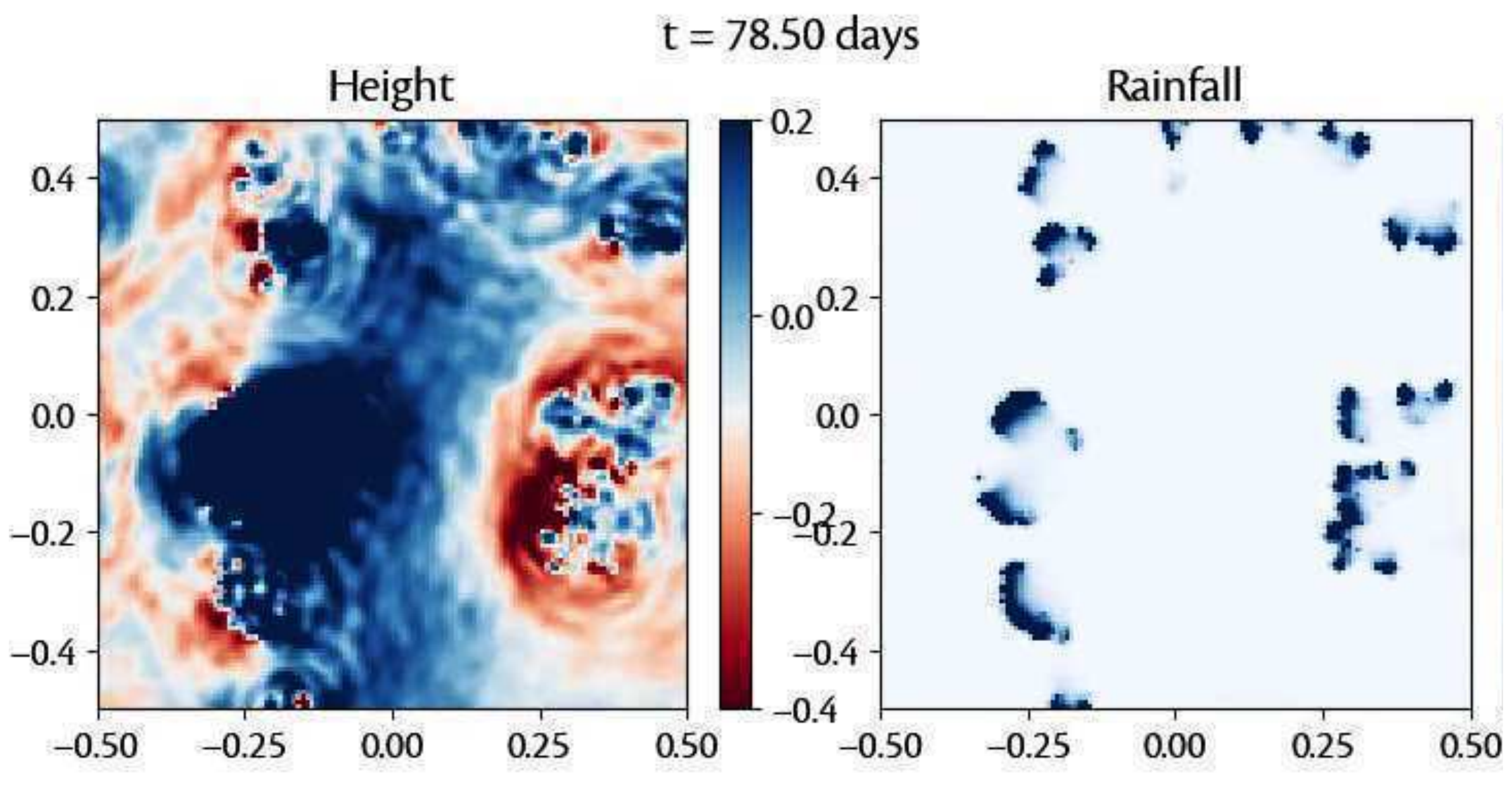} \\
     \caption{Snapshots of the height and precipitation fields at the times indicated, following an initial small perturbation at $y=0$ and $x = -0.3$ (Units of $x$ and $y$ are $10^7\m$.) The disturbance generates a front that propagates away from the disturbance before breaking up, but continuing in excitable, self-sustained motion. }
     \label{fig:evolve1}
 \end{figure}   
 }
 
 \subsection{Marginal excitability}

The boundary between the stable and excitable regimes, as determined numerically, is illustrated in \figref{fig:crit}.   To determine the boundary the equations are numerically solved in a domain of size $10^4\km$ in both \x- and \y-directions, periodic in the \textit{x}-direction and with a sponge near the meridional walls, and no rotation. (In later simulations with a nonzero beta the equator is in the center at $y=0$.)  In each case we choose $q_0 = q_g$, meaning that the state with no motion is just saturated and marginally unstable, by analogy to the tropical atmosphere \citep{Xu_Emanuel89}.  From that state, and for each set of parameters, a small perturbation is added and the system allowed to evolve freely. The system will either evolve into a self-sustained state or, after some time, return to its initial (unperturbed) state.  Depending on the result of a particular experiment the damping and/or rate of evaporation is increased or decreased until a marginally critical state is found, and each black dot in \figref{fig:crit} represents the convergence of such a process. There is an almost inverse relation between the critical values of the radiative damping and the evaporation rate; that is, higher values of radiative damping require a larger rate of evaporation to maintain excitability.   The other important moisture parameter is the latent heat of vaporization, $L_v$, and the dependence of kinetic energy, once the system reaches statistical equilibrium, is shown in \figref{fig:crit}c. The system is largely driven by the release of latent heat and consequently the kinetic energy increases approximately linearly with $L_v$.   Similar behavior occurs for a range of values of $\alphahat$ from 10 to 100, and here we use $\alpha = 60$, and there is no velocity dependence in the evaporation. 

The evolution from the small perturbation is characterized by an initial spike in the energy (\figref{fig:crit}b), very common in excitable systems (e.g., figure 7.1 of \cite{Izhikevich07}), followed either by a return to the initial state or by self-sustained motion.  A typical evolution in physical space from the initial state for an excitable state is shown in \figref{fig:evolve1}.  A small localized perturbation is applied that triggers convection,  and sends out a gravity wave,  leading to more convection and precipitation and a propagating, precipitating front. Initially the front is nearly circular but later breaks up into smaller fronts that in turn propagate with no preferred direction (in the absence of differential rotation) and decay, each one triggering convection nearby.  In the excitable regime the convection continues indefinitely, but in the stable regime the system slowly decays (typically on a timescale of tens of days), as in \figref{fig:crit}b.

\section{Effects of Rotation} 
\label{sec:nonlinear}
 We now explore the effects of rotation on the system when it is in an excitable parameter regime.  We first briefly discuss the case with constant rotation (i.e., flow on the $f$-plane)  and then, in more detail, the case on a beta-plane.

\subsection{Flow on the $f$-plane}
We apply a constant rotation of $f = 1\eten{-4}\ps$ and the perform experiments otherwise identical to those shown in \figref{fig:evolve1}. The initial evolution is very similar, but after a few days it is quite different, for the planetary rotation mitigates \textit{against} the organization of the convection. The reason is that the velocity induced by the condensational heat source tends to give rise to rotational motion around the heat source because of  the tendency toward geostrophic balance, rather than convergence toward it. (The effect is well illustrated by moving the heat source away from the equator in the Matsuno--Gill problem, as illustrated in figure 8.14 of \cite{Vallis17}.) This result should not be taken to mean that convective organization is impossible in the presence of rotation, since the omission of the nonlinear terms in the momentum equation means that shallow water cyclones do not form.

\subsection{Flow on a beta plane}
To  understand flow on a beta plane we first consider the case with humidity as a passive tracer, with no latent heat release, before considering the case where humidity feeds back on the flow.
 
\subsubsection{Humidity as a passive tracer}

\renewcommand*{\putab}{\put(-445,160){\small (a)}
\put(-215,160){\small (b) }
}
\begin{figure*}
   \centering
   % (a) \hspace{8cm} (b) \\
    \includegraphics[width=0.45\figwidth]{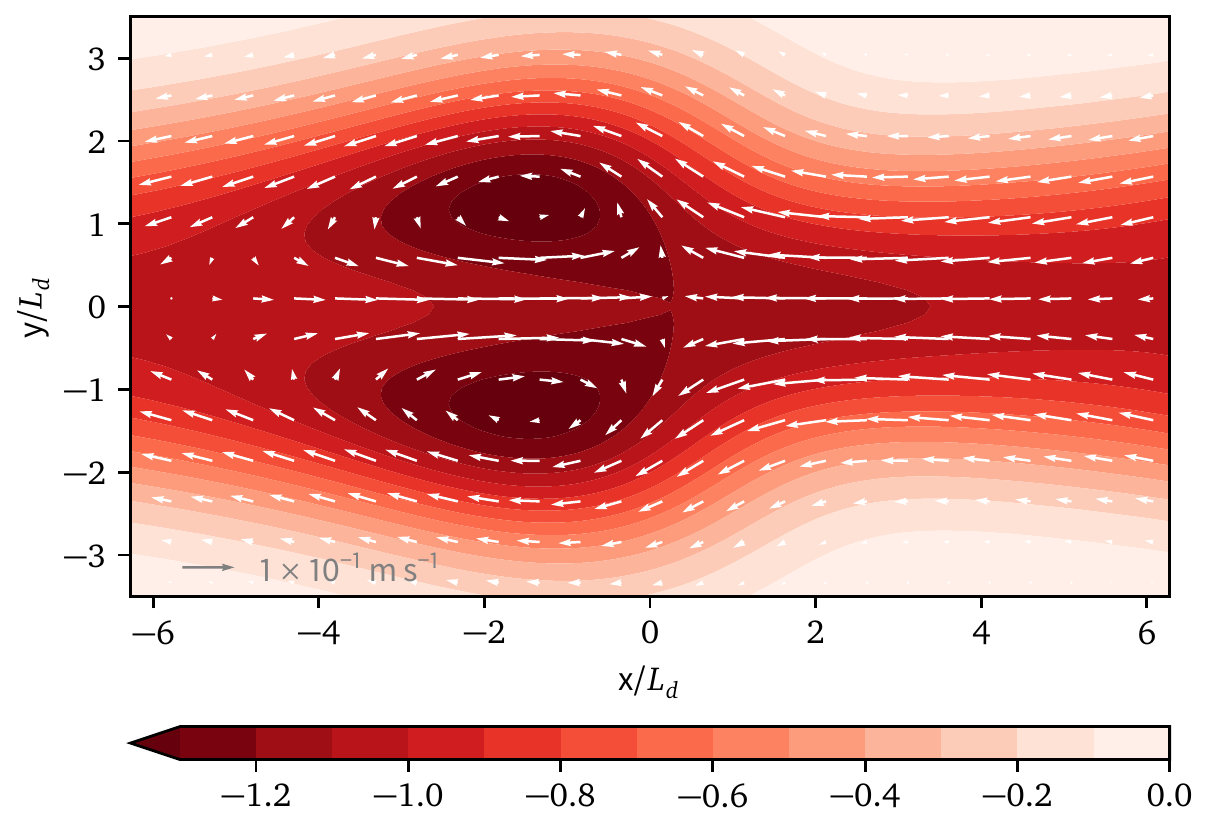}
     \includegraphics[width=0.45\figwidth]{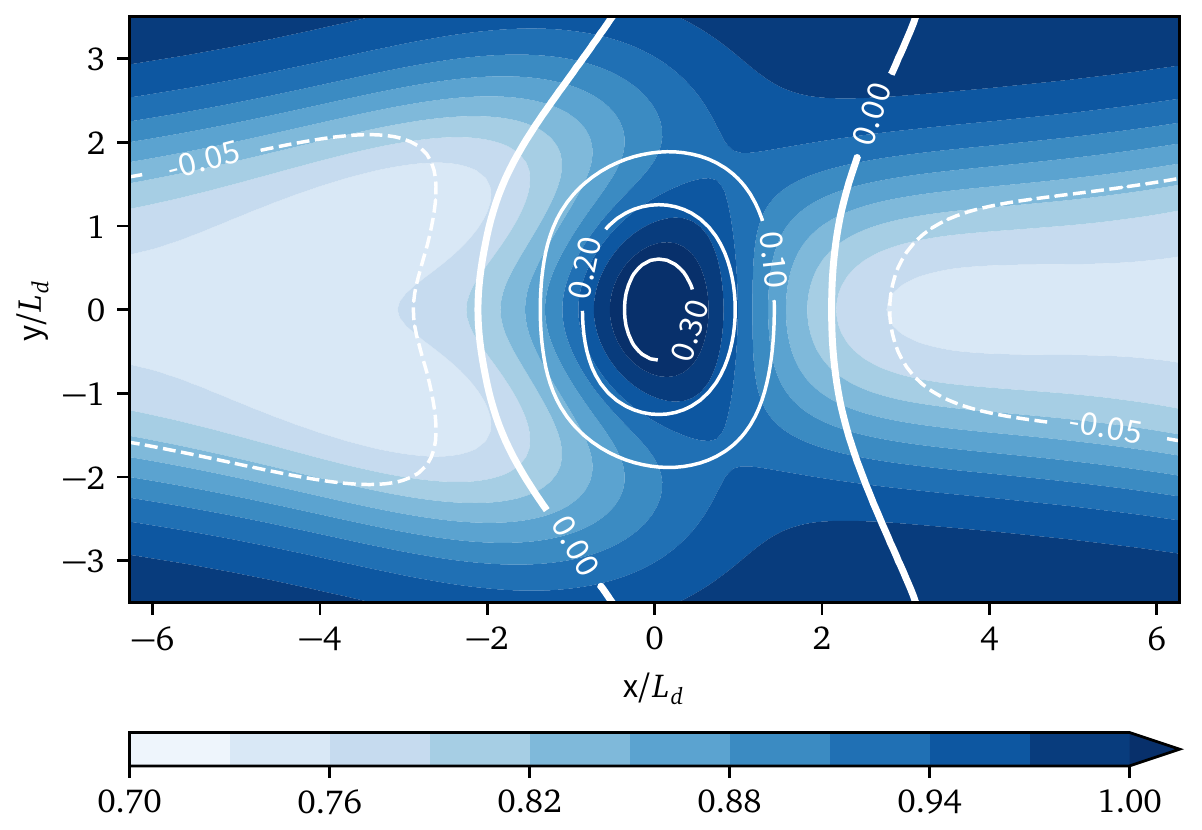}
     \putab
    \caption{Response to a steady heating centered at the origin on an equatorial beta-plane, where moisture is a passive tracer. (a) The height field (color filled contours) and wind vectors (arrows).  (b) The absolute humidity (specifically $q/q_0 - 1$, white contours) and relative humidity $q/\qsat$ (color filled contours), with darkest blue indicating saturation.}
    \label{fig:passive1}
\end{figure*}    

 We set $\gamma = 0$ in \eqref{eqp.1} but otherwise keep the equations unaltered, so that moisture evaporates from a saturated surface, is advected by the flow and condenses upon saturation. We integrate the equations on an equatorial beta-plane with $f = \fo + \beta y$, with $\fo = 0$ and $\beta = 2\eten{-11}\\m^{-1}\ps$.  The flow is forced by a static heating in the center of the domain of the form
\begin{equation}
	\label{ns.1} 
   	Q = h_0 \exp\left( - \frac{x^2 + y^2}{L_d^2/2} \right). 
\end{equation}
where $h_0$ is a constant.  The flow itself organizes into a familiar, steady,  \MG-like pattern, illustrated in \figref{fig:passive1}, with convergence very near to the center of the heating, depending slightly on the drag and relaxation parameters.  Whereas moisture accumulates near the center of the heating, it is the relative humidity that determines whether condensation occurs and this is a function of the height field as well as moisture. The large-scale warm anomalies of the Rossby lobes west and slightly poleward of the heating, and the  Kelvin waves east of the heating, inhibit precipitation except in regions close to the heating. 

\afterpage{\clearpage
\begin{figure}[H] \centering
    \includegraphics[width=1.\colwidth]{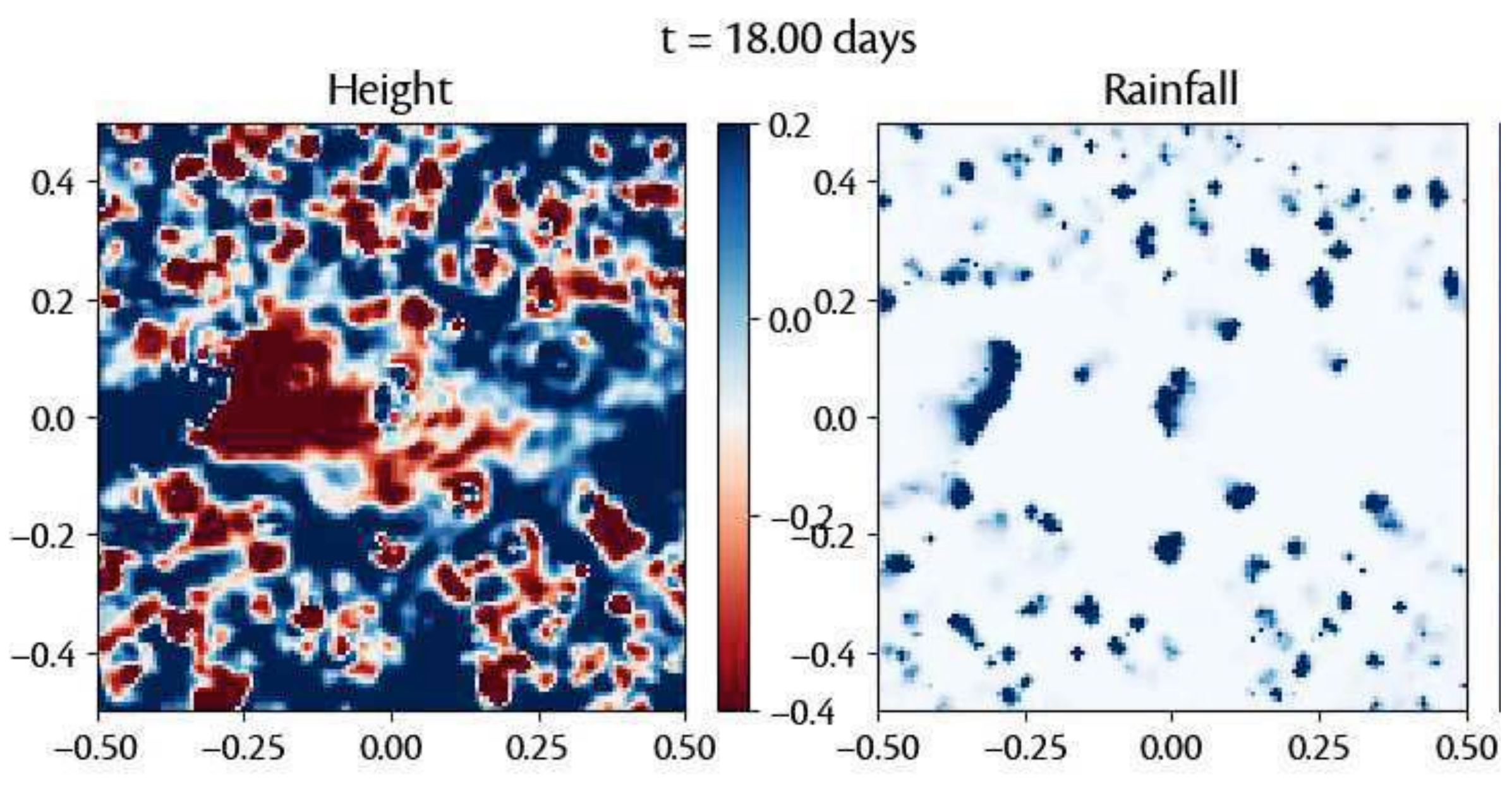} \\
    \includegraphics[width=1.\colwidth]{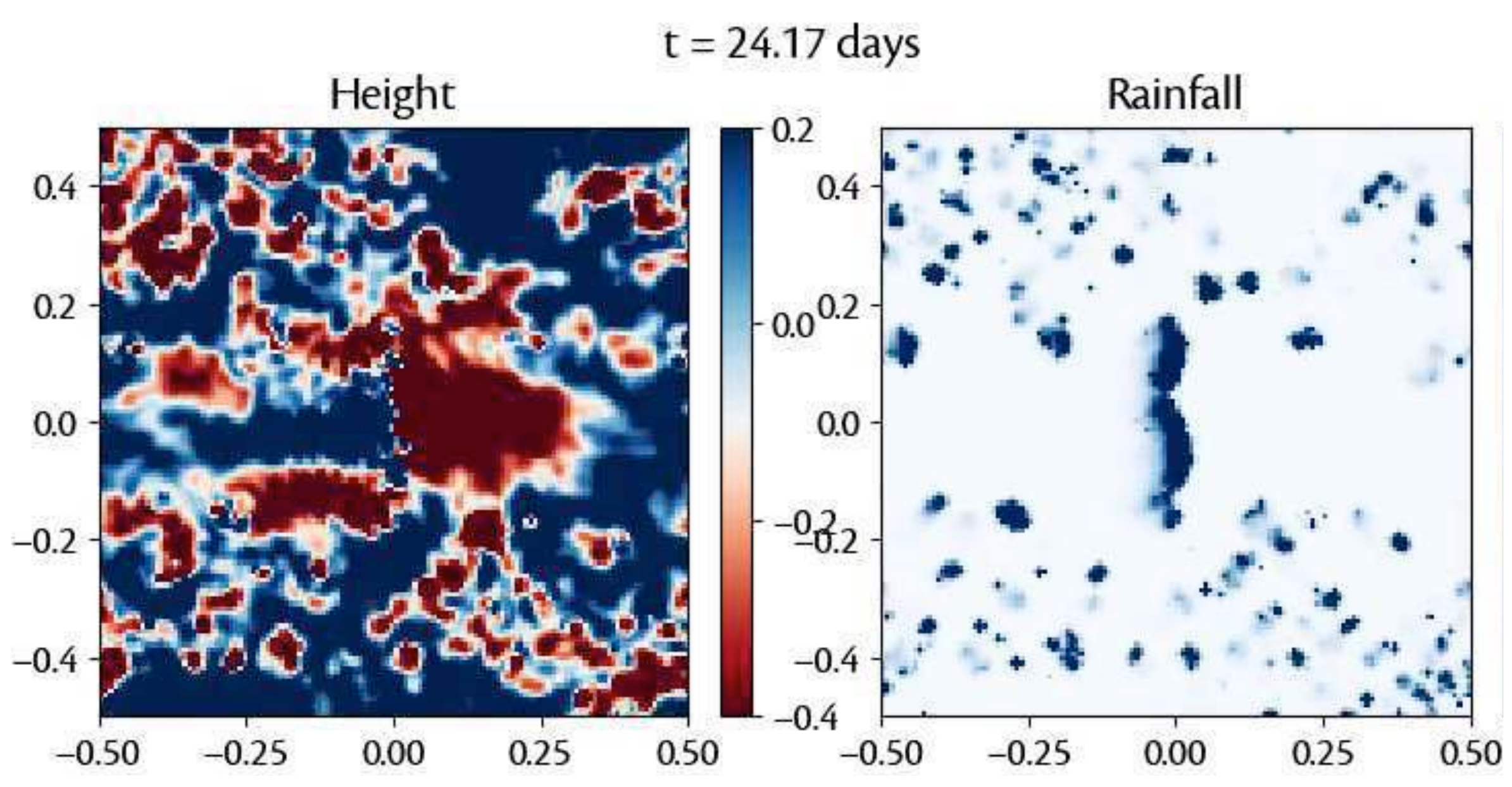} \\
    \includegraphics[width=1.\colwidth]{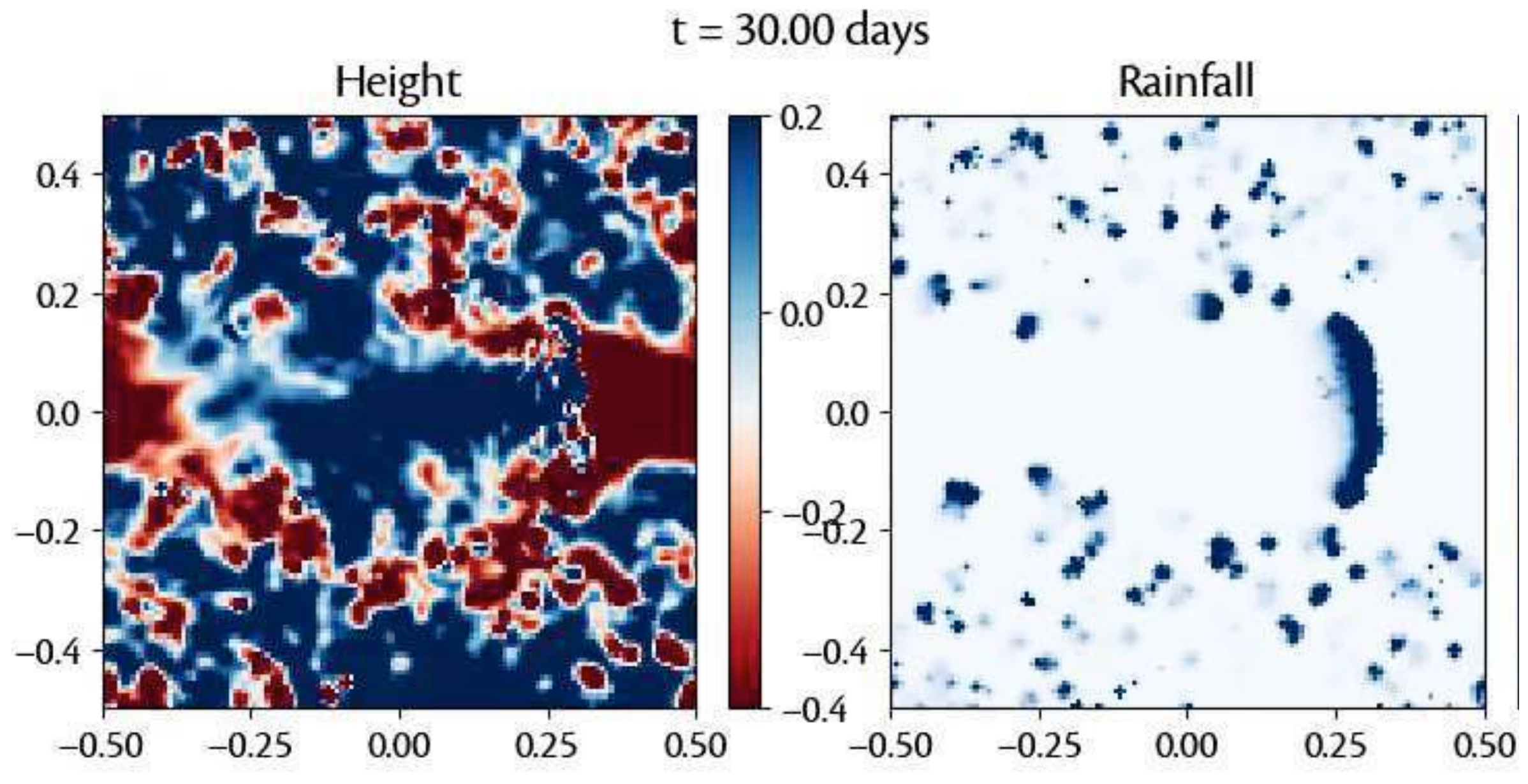} \\
    \caption{Snapshots of the height and precipitation fields at the times indicated, in simulations on a beta plane. The main disturbance forms about 18 days after initialization and propagates eastward at about 6\mps. Units of $x$ and $y$ are $10^7\m$ and the equatorial deformation radius is about $10^6\m$.}
    \label{fig:evolve2}
\end{figure}
}

Although there is a relative drying out of the equatorial region far to the east of the forcing, in the direct vicinity of the forcing the peak of relative humidity maximum is actually slightly to the east of the forcing center because of the asymmetric temperature distribution in the \MG pattern. This effect provides a mechanism for the eastward propagation of the pattern: the moisture convergence gives rise to condensation slightly east of the original heat source, so providing a new heat source. This in turn creates a pattern slightly east of the original one, and so on, thus providing a mechanism for eastward propagation. We now explore this further in a model with an active moisture tracer.

\subsubsection{Humidity as an active tracer}

On a beta-plane the  initial evolution from a localized perturbation is similar to that in the non-rotating case shown in \figref{fig:evolve1}.  At subsequent times rotational effects inhibit the aggregation of convection in mid-latitudes, but close to the equator the convection organizes itself. Condensation near the equator provides a heat source, which tries to generate a pattern similar to that of the left panel of \figref{fig:passive1}, with moisture convergence maintaining the heat source.  Unlike the case with no rotation, there is an east-west asymmetry, for two related reasons. First, the Matsuno--Gill pattern itself is asymmetric, bringing warm, moist air from the east.  The associated convergence leads to condensation on the leading (i.e., eastern) edge of the existing condensational heat source, leading to the formation of a precipitating front, related to those described by \cite{Frierson_etal04} and \citet{Lambaerts_etal11}.  Second, close to the equator Kelvin waves are excited and these propagate east, triggering convection and generating more convection in the moist converging fluid just east of the initial disturbance, and so on.  If the system is in an excitable regime the mechanism is self-sustaining and the precipitation front propagates eastward. A typical progression is shown in \figref{fig:evolve2}; this simulation has $\alpha=60$, $\gamma = 15$, $\tau_r = 2 \,\text{days}$, $\lambda = 0.08\, \text{days}^{-1}$ and $\beta = 2\eten{-11}\m^{-1}\ps$.  Simulations with resolutions from 100\km to 25\km show very similar behavior. The associated height pattern is qualitatively similar to a Matsuno--Gill pattern with a warm, leading Kelvin lobe eastward of the precipitating confined to the equator and Rossby lobes flanking and slightly westward of the precipitation, broadly consistent with the composite patterns seen in observations \citep{Kiladis_etal05, Adames_Wallace14}. At some times (e.g. at 30 days) the Kelvin lobe itself is flanked by geopotential disturbance of opposite sign, giving a quadrupole nature to the height field. The latitudinal width of the disturbance is of order the equatorial deformation radius, $\sqrt{c/\beta}$, where $c = \sqrt{gH}$, and which in the simulations shown is about 1000\km. For latitudes beyond this distance from the equator the convergence is too weak to generate an organized pattern, as in the $f$-plane simulations.

\begin{figure}
\centering
    \includegraphics[width=0.85\colwidth]{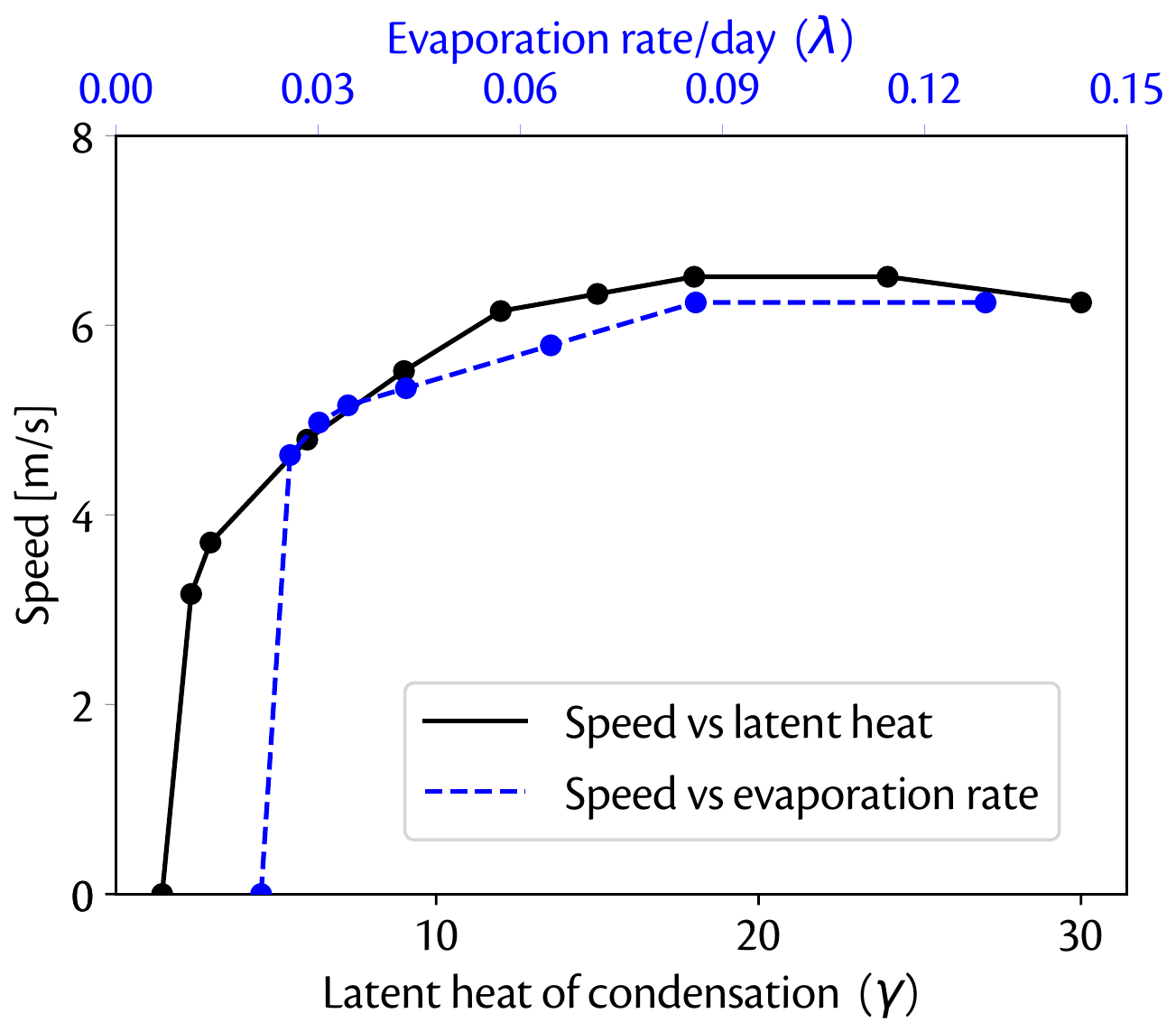} 
    \caption{Speed of the main, MJO-like, eastward moving disturbance as function of the evaporation coefficient $\lambda$ and  the latent heat of condensation, $\gamma$. Where $\gamma$ varies the value of $\lambda$ is fixed at 0.08 and when $\lambda$ varies the value of $\gamma$ is 15.}
    \label{fig:speed}
\end{figure}

The speed of the disturbance is not directly associated with the dry gravity wave speed, or even a moisture-modified gravity wave.  Rather, it is associated with the time taken for the circulation to respond to a heat source: the disturbance cannot move so quickly that the Matsuno--Gill-like pattern cannot keep up with it and maintain the supply of water vapor to it.  The strength of the pattern, and thus its timescale, is determined by the release of latent heat, and as \figref{fig:speed} shows the speed increases as either the latent heat of condensation increases or as the efficiency of evaporation increases. The speed does not increase without bound, for as the effects of moisture increase  the pattern formation becomes more irregular, and two or more precipitating patterns may form along the equator, each stealing moisture from the other and slowing the propagation.

Numerical experiments demonstrate that the robustness of the formation of a coherent equatorial structure and its eastward propagation is enhanced by the presence of a wind-evaporation feedback \citep{Neelin_etal87, Emanuel_etal87}. Specifically, suppose evaporation, $E$, is parameterized via a term of the form
\begin{equation}
	\label{wishe.1} 
   	E = \lambda {|\ub + \bm{U}| \over U_0} (q_0 - q),
\end{equation}
where $\ub$ is the model produced wind and $\bm{U}$ is a constant, background wind due to, say, the trades. If $\bm U$ is directed westward, then evaporation is enhanced eastward of the disturbance and, if $|\bm{U}| \approx 1\mps$ (similar to wind speeds produced internally by the disturbance itself) the formation of an eastward propagating disturbance is enhanced. If $\bm U$ is eastward with a similar magnitude then MJO-like disturbances are eliminated.  If simulations are performed with a nonzero value of $\bm U$ with no background rotation ($f_0 = \beta = 0$) then the fronts that form preferentially move eastward or westward, depending on whether $\bm U$ is directed westward or eastward, respectively.   

A well-known feature of the MJO is that it forms over the warm waters of the Indian Ocean, and that the next event formation will occur some 30--60 days later. It is, in fact, a common feature of an excitable system that it cannot support the passage of a second disturbance over a given location until sufficient time -- the `refractory period' -- has passed \citep{Izhikevich07}, and this is also a property of our model.  To demonstrate this we impose a spatially varying distribution of surface humidity, as one would find with a varying sea-surface temperature anomaly, as in the left panel of \figref{fig:position}. The right panel shows the position of the main disturbance, as determined by the location of the maximum of the precipitation divergence in the velocity field. Disturbances typically form only in the vicinity of the maximum surface humidity and propagate east, decaying where the surface humidity is too low. A second disturbance cannot form until the first disturbance is sufficiently distant from the genesis location, because the first disturbance leaves a wake of dry air that needs sufficient time to reform into a converging pattern and initiate the moisture feedback. In the simulation shown  the reformulation time is roughly 20-40 days, the timescale associated with  the decay of the disturbance as it moves east.  Even without a previous MJO disturbance the atmosphere typically takes many days to organize itself into a quasi-steady state in response to a localized heating \citep{Heckley_Gill84}.

\newcommand*{\puta}{\put(-225,160){\small (a)} }
\newcommand*{\putb}{\put(-225,160){\small (b)} }
\afterpage{
\begin{figure}[H] 
     (a)  \hspace*{7cm}  (b) \\
     \includegraphics[width=0.6\colwidth]{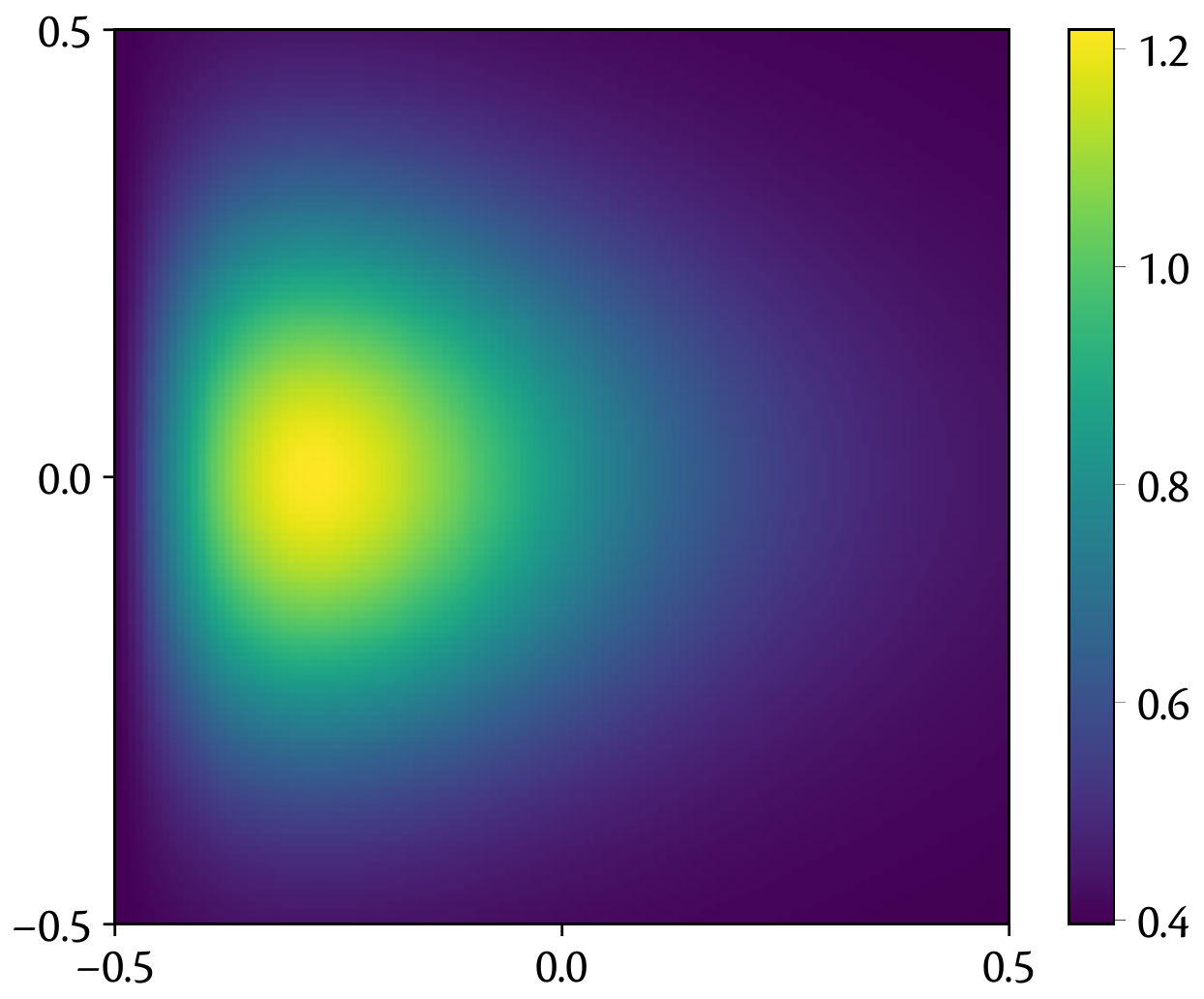}  \qquad 
     \includegraphics[width=0.85\colwidth]{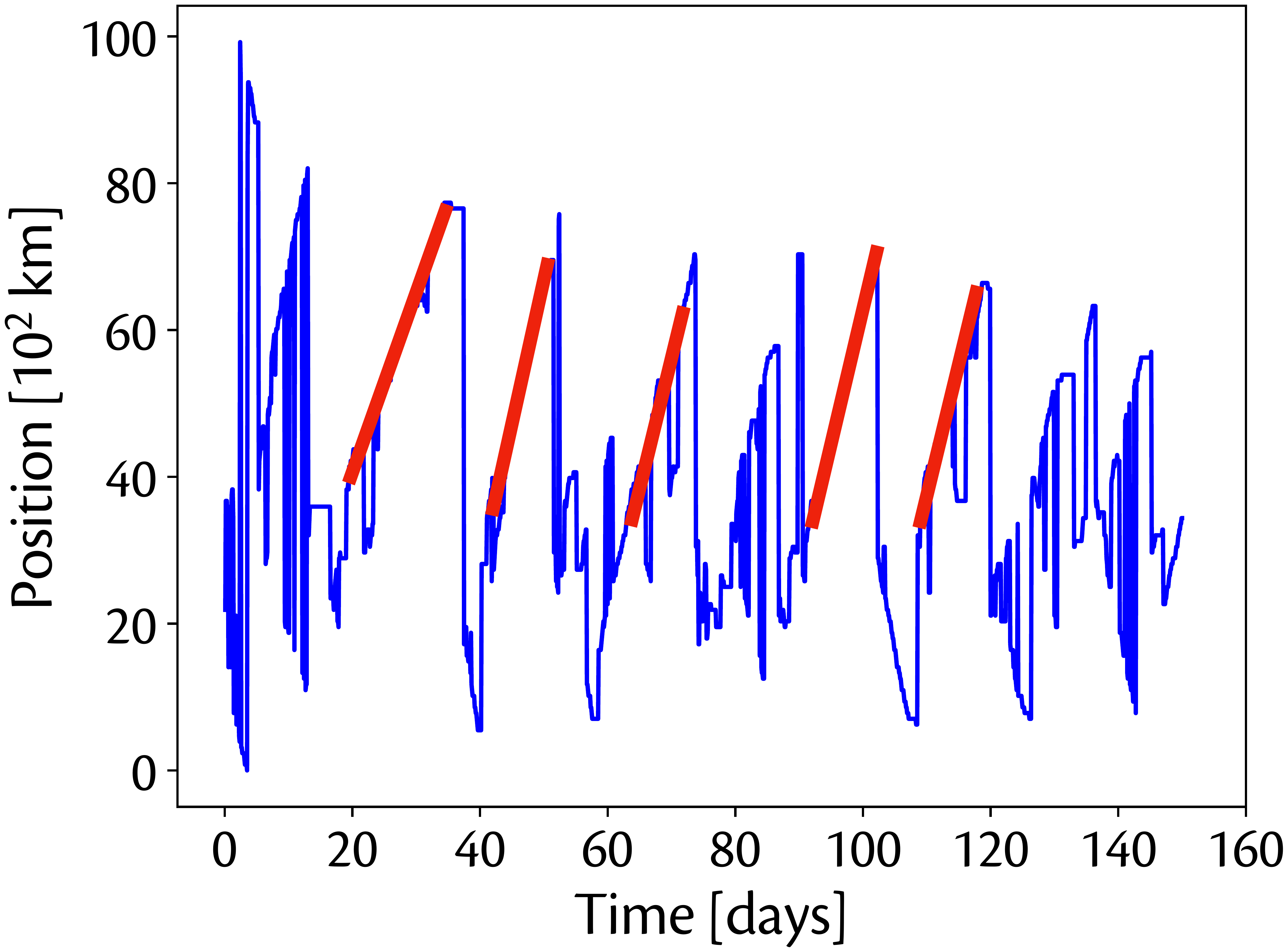} 
     \caption{(a) Sea-surface humidity, representing a warm pool in the western equatorial region with a fraction humidity about 25\% above a base level  (b) Position of the maximum precipitation at the equator, with thicker red lines indicating where the precipitation is associated with MJO-like activity. MJO events form only in the warm western region and decay before reaching the eastern edge, and here recur every 20--40 days.}
     \label{fig:position}
\end{figure}
}

\section{Discussion and Conclusions} \label{sec:discussion}

We have presented a simple, explicit model that reproduces many of the main features of intra-seasonal variability and the Madden--Julian oscillation in the tropical atmosphere, and described the mechanism that causes these features.  By simple and explicit we mean that we treat the atmosphere as a single baroclinic mode in the vertical and we do not use any convective parameterization (except a rather basic one), or make additional approximations about the scale and nature of the system; rather, we directly solve the resulting equations of motion, which consist simply of the shallow water equations plus a humidity variable.   Condensational heat release affects the height field (a proxy for temperature), which induces a velocity field, leading to more convection and so on.  If the condensational effects are sufficiently strong, as determined by the latent heat of evaporation and the efficiency of evaporation from the surface, and the radiative damping appropriately weak, the system is excitable and self-sustained motion ensues. Excitable behavior occurs over a wide range of physically reasonable parameters and in this regime the motion -- which can develop quite large scales --  is driven by the condensational heating at small scales, without the need for any large-scale instabilities. 

\begin{figure}
 \centering
     \includegraphics[width=0.8\textwidth]{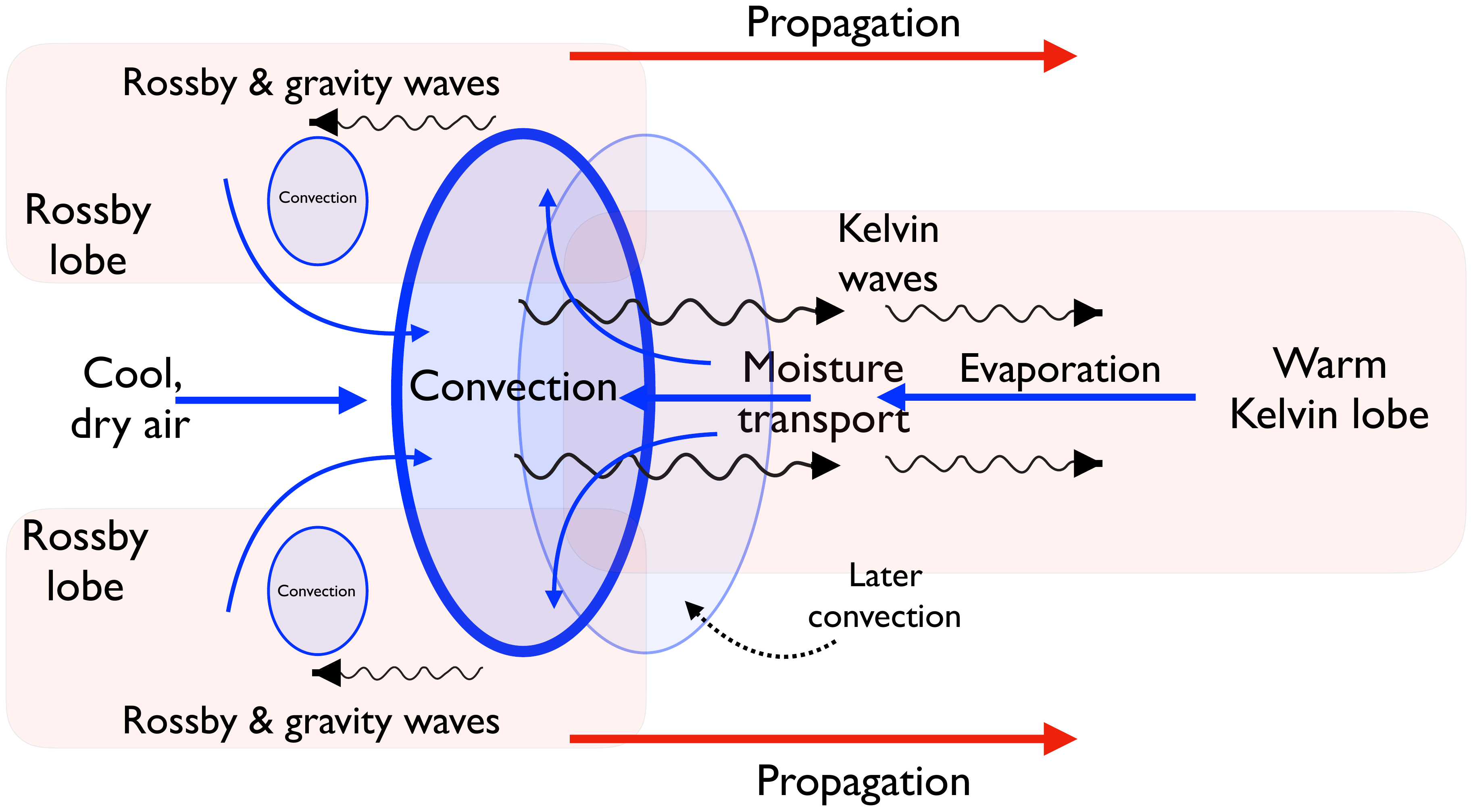} 
     \caption{Schematic of an eastward propagating equatorial disturbance. Convection at the equator gives rise to a feedback producing convective aggregation and a modified Matsuno--Gill-like pattern. Warm moist air is then drawn in from the east but this is convectively unstable, amenable to triggering by the eastward propagating gravity waves from initial disturbance, and new convection forms on the eastern edge of the original location. The whole pattern then moves unsteadily eastward.}
     \label{fig:schematic}
\end{figure}

When integrated on the beta plane the motion in the tropics becomes organized. In a background state that is marginally stable convection aggregates around the equator creating a Matsuno--Gill-like pattern with a  scale determined by the equatorial deformation radius.  However, the pattern is unstable in the sense that the  combination of moist convergence and gravity waves triggers convection nearby and the disturbance propagates,  preferentially in an eastward direction. The directionality arises because the pattern draws warm moist air in from the east along the equator, and this air becomes conditionally unstable. Convection is triggered at the eastern edge of the existing disturbance and the whole system then propagates east, as sketched in \figref{fig:schematic}. Note that the moisture is not advected eastward in the disturbance; rather, it has an evaporative source and is drawn in \textit{from} the east.  The large-scale disturbance may be surrounded by a collection of smaller-scale convective events, and these have no preferred direction because the pressure field they induce is more nearly isotropic. 

Although simple in construct, the model displays some of the key observed features of the Madden--Julian oscillation, notably:
\begin{enumerate}
    \item A predominantly dipolar (although  at times quadrupolar) Matsuno--Gill-like pattern in the height field, with a Kelvin lobe extending east, flanked by two off-equatorial Rossby lobes. The size of the structure depends on the equatorial radius of deformation.
    \item The whole pattern moves eastward at a speed of a few meters per second. The speed is largely determined by the levels of moisture in the system and by the magnitude of the latent heat of condensation, which determine the strength of the pattern and the time it can take to form and move itself.
    \item The pattern preferentially forms where there is greater availability of moisture, corresponding to a higher sea surface temperature. If a pattern forms and moves east from that location then an interval of order a few tens of days must pass before a second disturbance can form in the dry wake of the first.
\end{enumerate}    
None of these features are built-in to the model; rather, they are all emergent properties. The presence of a wind-induced evaporation, or  WISHE, also has an noticeable effect on its eastward propagation, as in \cite{Khairoutdinov_Emanuel18}. Since the energy source is, ultimately, evaporation from the surface,  a higher SST (i.e., a higher level of surface humidity) produces a more energetic simulation, as in \cite{Arnold_etal15}, and is responsible for the localization of the genesis region.  A rather less realistic feature of the model is that the moisture convergence and feedback onto the height field lead to the formation of fronts that are too narrow in the along-propagation direction, compared to observations of composite MJOs. However, both observations from TRMM and high resolution three-dimensional simulations do show that the region of intense precipitation tends to become meridionally extended and zonally confined as the MJO enters the western Pacific \citep{Liu_etal09}.  It is not surprising that the model cannot capture the details of the precipitation distribution but the larger scale features appear broadly consistent with observations.  Tight frontal formation is in fact commonly found in idealized moist systems \citep{Yano_etal95, Frierson_etal04,  Lambaerts_etal11}, and the gridpoint generation of storms and fronts are a common feature of moist GCMs without a convective parameterization. 

The model also suggests why many GCMs are unable to produce MJO-like variability, even as some cloud permitting models with no convective parameterization but at similar resolution are able to do so \citep{Khairoutdinov_Emanuel18}. The mechanism itself is not especially sensitive to resolution, but it does require that the system be excitable. Now, cloud-resolving models in a statistical radiative-convective equilibrium over a uniform SST are almost certainly excitable systems (even if not normally described as such), but a convective parameterization that  distributes convective effects more smoothly will perforce diminish the gravity wave generation and that excitability.  Certainly, the nature of the convective parameterization \textit{does} affect the production of an MJO in a GCM \citep{Benedict_etal14}, and the explicit introduction of high frequency variability may be beneficial, as in \cite{Deng_etal15}.  Relatedly, many GCMs are unable to produce the quasi-biennial oscillation (QBO) without a gravity wave drag parameterization, in part because the upwardly propagating Kelvin and Rossby waves are not properly simulated.  The difference is that in the QBO the gravity waves propagate up toward the maximum zonal wind, drawing its maximum down, whereas in the MJO the disturbance itself is the source of the gravity waves and the flow convergence, moving the system east.

The interaction of the moisture field with the pressure and wind fields is essential to the mechanism we have presented here, as it is in the moisture-mode theories discussed in the introduction.  The MJO we envision is not, however, that of a single large-scale coupled mode; rather, the large-scale structure and propagation arise because the tropical atmosphere is an unstable system living on the beta-plane.  Travelling disturbances and pattern formation are common in excitable systems \citep{Meron92} and moist radiative-convective systems appear to be no exception.  The beta-plane dynamics add an extra twist and lead to aggregation at the equator, with warm, moist air drawn from the east toward the disturbance. New convection is then triggered on its eastern flank and the whole pattern bootstraps its way east.

The mechanisms described above seem consistent with the evolution occurring in some more comprehensive 3-D models \citep{Nasuno_etal09, Arnold_Randall15, Khairoutdinov_Emanuel18}, albeit with differences in structure and detail.  Nevertheless, the respective importance of the processes at work in the various models is unclear and achieving a better understanding of the connection between them, and the real atmosphere, remains a topic for future work. The ENSO system may be another example of an excitable geophysical system, with a refractory period of a few years and excited by westerly wind bursts, but that too remains to be investigated.

\subsubsection*{Acknowledgements}
This work was funded by the Leverhulme Trust,  NERC and the Newton Fund. I thank Professor Pete Ashwin for a very useful conversation about excitable systems and Professors Mat Collins and John Thuburn for comments about the MJO and possible ENSO relevance.

~ \\
\footnotesize The End.

\end{document}